\documentclass[pre,aps,twocolumn,superscriptaddress]{revtex4}

\usepackage[dvips]{graphicx}
\usepackage{amssymb,amsfonts,amsmath}
\usepackage{color}
\usepackage{ulem}
\usepackage[hidelinks]{hyperref}
\usepackage{bbm}
\usepackage{bm}

\newcommand{\rv}{{\mathbf r}}

\newcommand{\Tr}{{\rm Tr}\,}
\newcommand{\e}{{\rm e}}

\newcommand{\pv}{{\bf p}}

\newcommand{\Fv}{{\bf F}}

\newcommand{\eps}{{\boldsymbol \epsilon}}
\newcommand{\unity}{{\mathbbm 1}}
\newcommand{\cov}{{\rm cov}}

\newcommand{\avg}[1]{\Big\langle {\protect #1} \Big\rangle}
\newcommand{\eqr}[1]{Eq.~\eqref{#1}}

\newcommand{\mydelete}[1]{{}}
\newcommand{\taub}{{\boldsymbol\tau}}

\newcommand{\rmint}{{\rm int}}
\newcommand{\rmext}{{\rm ext}}
\newcommand{\rmeq}{{\rm eq}}
\newcommand{\commaz}{}

\begin{document}

\title{Hyperforce Balance via Thermal Noether Invariance of Any Observable}

\author{Silas Robitschko}
  \affiliation{Theoretische Physik II, Physikalisches Institut, 
    Universit{\"a}t Bayreuth, D-95447 Bayreuth, Germany}
\author{Florian Samm\"uller}
\affiliation{Theoretische Physik II, Physikalisches Institut, 
  Universit{\"a}t Bayreuth, D-95447 Bayreuth, Germany}
\author{Matthias Schmidt}
\affiliation{Theoretische Physik II, Physikalisches Institut, 
  Universit{\"a}t Bayreuth, D-95447 Bayreuth, Germany}
\email{Matthias.Schmidt@uni-bayreuth.de}
\author{Sophie Hermann}
\affiliation{Theoretische Physik II, Physikalisches Institut, 
  Universit{\"a}t Bayreuth, D-95447 Bayreuth, Germany}

\date{4 September,
revised version: 20 December 2023,  second revision: 7 February 2024}

\begin{abstract}
Noether invariance in statistical mechanics provides fundamental
connections between the symmetries of a physical system and its
conservation laws and sum rules. The latter are exact identities that
involve statistically averaged forces and force correlations and they
are derived from statistical mechanical functionals. However, the
implications for more general observables and order parameters are
unclear. Here, we demonstrate that thermally averaged classical phase
space functions are associated with exact hyperforce sum rules that
follow from translational Noether invariance. Both global and locally
resolved identities hold and they relate the mean gradient of a
phase-space function to its negative mean product with the total
force. Similar to Hirschfelder's hypervirial theorem, the hyperforce
sum rules apply to arbitrary observables in equilibrium. Exact
hierarchies of higher-order sum rules follow iteratively. As
applications we investigate via computer simulations the emerging
one-body force fluctuation profiles in confined liquids. These local
correlators quantify spatially inhomogeneous self-organization and
their measurement allows for the development of stringent convergence
tests and enhanced sampling schemes in complex systems.
\end{abstract}

\maketitle

\section{Introduction}
The task of predicting thermal averages of phase space functions lies
at the center of attention in Statistical Mechanics. Prominent
examples include correlation functions and order parameters, but also
global quantities such as internal and external energies, the entropy,
and much more are considered \cite{hansen2013,evans1979}. Significant
progress has been reported for problem-specific order parameters that
are tailored to capture intricate correlations effects. Recent
examples that address the spatial ordering behaviour of dense liquids
include beyond-two-body correlation functions, as advocated by Kob and
coworkers \cite{zhang2020pnas,singh2023pnas} and by Janssen and her
coworkers \cite{pihlajamaa2023}.

In constrast to such freedom of choice, the variables within classical
density functional theory \cite{evans1979, evans1992, evans2016} seem
to be a priori uniquely determined by the existence of a generating
free energy functional and the associated structure of pairs of
conjugate variables, which in particular are the external one-body
potential energy $V_\rmext(\rv)$ and the density profile
$\rho(\rv)$. However, there are recent extensions to density
functional theory to systematically include the local compressibility
\cite{evans2015jpcm, evans2019pnas, coe2022prl}, which forms a
well-accessible order parameter for local particle number
fluctuations.  Technically, the local compressibility constitutes
either a parametric derivative of the equilibrium density profile with
respect to the chemical potential or, analogously, the covariance of
the local density and the global particle number. A generalization
from such chemical particle number fluctuation to thermal fluctuations
has been recently performed
\cite{eckert2020,eckert2023fluctuation}. Working in the grand
ensemble, where the particle number fluctuates, is thereby crucial to
not impose artificial constraints on the system.

Besides the standard thermodynamical thinking in terms of thermal and
chemical equilibrium, there is much recent progress in the force point
of view. Highly efficient force sampling techniques allow to obtain
reliable results within many-body simulations that outperform more
straightforward counting methods \cite{rotenberg2020, borgis2013,
  delasheras2018forceSampling, purohit2019, coles2019, coles2021,
  coles2023revelsMD}. Forces are also at the core of power functional
theory \cite{schmidt2022rmp} as a systematic approach to formulate
coupled many-body dynamics on the one-body level of dynamical
correlation functions.

To be specific, in thermal equilibrium of a spatially inhomogeneous
system, the sum of all mean forces necessarily vanishes at each
position $\rv$. This is expressed by the following exact sum rule:
\begin{align}
  \Fv_\rmint(\rv) 
  - \rho(\rv) \nabla V_\rmext(\rv) 
  &= k_BT \nabla\rho(\rv).
  \label{EQybg}
\end{align}
Here $\Fv_\rmint(\rv)$ is the localized force density that acts at
position $\rv$ due to the interparticle interactions with all
surrounding particles, $\nabla$ denotes the derivative with respect to
$\rv$ such that $-\nabla V_\rmext(\rv)$ is the external force field,
$k_B$ indicates the Boltzmann constant, and $T$ is absolute
temperature.

The sum of the interparticle and external force densities on the left
hand side of \eqr{EQybg} balances the thermal diffusive contribution
on its right hand side. This is a classical result due to Yvon, Born
and Green (YBG)~\cite{hansen2013}, where for particles that mutually
interact only via a pair potential, the interparticle force density
$\Fv_\rmint(\rv)$ is expressed as an integral over the two-body
density multiplied by the pair force \cite{hansen2013}.  Higher-order
versions of \eqr{EQybg} form a hierarchy. That the first YBG equation
\eqref{EQybg} has practical consequences for carrying out sampling
tasks in simulations is only a quite recent insight. Loosely speaking,
{\it force sampling} \cite{rotenberg2020, borgis2013,
  delasheras2018forceSampling} amounts to obtaining simulation data
for the left hand side of \eqr{EQybg} and then in a post processing
step dividing by $k_BT$ and building the inverse operation of the
spatial derivative on the right hand side via suitable integration in
position. This method yields results for the density profile
$\rho(\rv)$ which feature a significant reduction of statistical
noise~\cite{borgis2013, rotenberg2020, delasheras2018forceSampling,
  purohit2019, coles2019, coles2021, coles2023revelsMD}.

Exploiting Noether's Theorem \cite{noether1918, byers1998} in
statistical mechanics has been performed in a variety of ways
\cite{baez2013markov, marvian2014quantum, sasa2016, sasa2019,
  revzen1970, budkov2022, brandyshev2023, bravetti2023}.  Considering
the invariance of statistical mechanical functionals leads very
naturally to the notion of statistically averaged forces when spatial
displacement is imposed on the system; mean torques emerge when
invariance against rotations is addressed \cite{hermann2021noether,
  hermann2022topicalReview}. In previous work, we have shown that the
statistical Noether concept also applies quantum mechanically
\cite{hermann2022quantum} and that it gives access to global force
fluctutations \cite{hermann2022variance}. Generalizing to local
invariance \cite{hermann2022quantum, tschopp2022forceDFT} that is
resolved in spatial position facilitated fresh insights into the
correlation structure of the liquid state
\cite{sammueller2023whatIsLiquid}. Considering the first order in the
displacement field yields the thermal equilibrium force balance
relationship according to the~YBG equation \eqref{EQybg}
\cite{hermann2022quantum, tschopp2022forceDFT}. At second order
hitherto unkown two-body force-gradient and force-force correlators
emerge and these, together with the standard pair correlation
function, are constrained by exact Noether identities
\cite{sammueller2023whatIsLiquid}.

This situation of theory development leaves open the question of
whether more general observables that serve as important order
parameters and quantifiers of spatial structure will also be affected
by the statistical Noether invariance, as one could glean from the
generality of the thermal invariance concept. Here we demonstrate that
any statistical observable $\hat A$ is intrinsically associated with a
corresponding hierarchy of exact identities that emerges from its
statistical shifting invariance properties. We validate the
corresponding exact local and global sum rules for a range of relevant
observables via many-body simulations of a confined Lennard-Jones
fluid. The results clarify a very intimate link of global and locally
resolved correlators and they suggest a very general statistical
mechanical structure.

Our framework can be viewed as a generalization of the YBG
equation~\eqref{EQybg} to systematically include the dependence on a
further given observable $\hat A$. The equilibrium force balance
\eqref{EQybg} itself is recovered for the trivial case $\hat A=1$.
Such generalization is not uncommon in Statistical Mechanics. The
relationship of our theory and the~YBG force balance equation is akin
Hirschfelder's hypervirial theorem \cite{hirschfelder1960} as a
generalization of the standard virial theorem \cite{hansen2013} to
also invoke an additional dependence on a given phase space
function~$\hat A$.  Our theory can hence be viewed as a hyperforce
balance relationship, and we derive global and local variants below,
see Eqs.~\eqref{EQAepsGlobal} and~\eqref{EQAepsLocal}. We further show
that the local version simplifies further to
\eqr{EQAepsCoordinatesOnly} and more explicitly to
\eqr{EQAepsCoordinatesExplicitForm} in case of $\hat A$ being
independent of momenta; see Fig.~\ref{FIG6} for an overview of
results.  As a specific example, our methodology not only allows to
sample density gradients, as is possible in force sampling schemes
\cite{borgis2013, rotenberg2020, delasheras2018forceSampling,
  purohit2019, coles2019, coles2021, coles2023revelsMD}, but also to
sample force density gradients.  The general method complements
existing counting and force-sampling techniques and it gives much
inspiration for rigorous statistical mechanical theories based on
exact identities.  As we lay out, the degree of numerical accuracy to
which the Noether sum rules are satisfied can serve as an estimator
for sufficient equilibration of slowly converging systems.

The paper is organized as follows. Starting from the description of
the underlying Statistical Mechanics, we present the general Noether
invariance theory in Sec.~\ref{SECtheory} for cases of global and
locally resolved shifting invariance. The theory is applied to
specific observables, tested in a confined fluid, and its implications
for sampling methodology are discussed in Sec.~\ref{SECapplications}.
We give conclusions and an outlook in Sec.~\ref{SECconclusions}.

\section{Methods}
\label{SECtheory}

\subsection{Statistical Mechanics}
We consider general thermal many-body systems of particles with
identical mass $m$, coordinates $\rv_1,\ldots,\rv_N\equiv \rv^N$, and
momenta $\pv_1,\ldots,\pv_N\equiv \pv^N$, where $N$ denotes the number
of particles. The Hamiltonian is of the standard form $H=\sum_i
\pv_i^2/(2m) + u(\rv^N) + \sum_i V_\rmext(\rv_i)$, where the sums
$i=1,\ldots,N$ run over all particles, $u(\rv^N)$ is the interparticle
interaction potential, and $V_\rmext(\rv)$ is an external potential
that depends on position~$\rv$. Thermal equilibrium is characterized
by a statistical equilibrium ensemble with grand canonical probability
distribution $\Psi_{\rm eq}=\Xi^{-1}\e^{-\beta (H-\mu N)}$, where
$\beta=1/(k_BT)$ and $\mu$ indicates the chemical potential. The
normalization factor of $\Psi_\rmeq$ is the partition sum $\Xi=\Tr
\e^{-\beta (H-\mu N)}$, where the classical trace is defined as
$\Tr\cdot=\sum_{N=0}^\infty (N!  h^{3N})^{-1} \int d\rv^N\int d\pv^N
\cdot$, with~$h$ indicating Planck's constant. The thermal equilibrium
average $A$ of a given phase space function $\hat A$ is then obtained
as $A = \langle \hat A \rangle \equiv \Tr \Psi_{\rm eq} \hat A$, where
we have suppressed the dependence of $\hat A$ on the phase space
variables~$\rv^N$ and $\pv^N$ in the notation, i.e., in full notation
we have $\hat A(\rv^N,\pv^N)$ as well as potentially further
parametric dependence such as on a generic position variable $\rv$.

\begin{figure}[!t]
\includegraphics[page=1,width=.99\columnwidth]{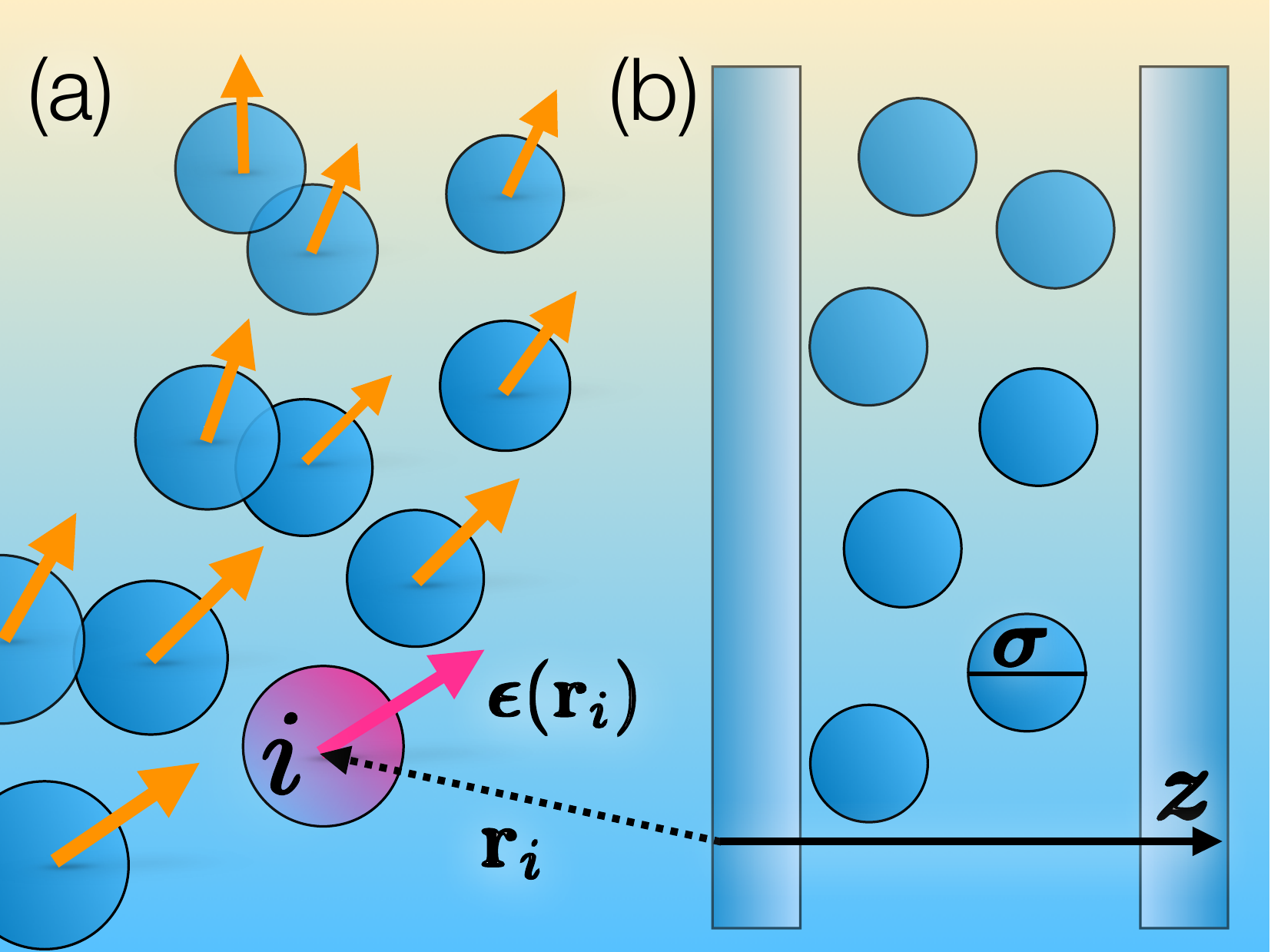}
  \caption{Illustrations of the relevant geometries. (a) The shifting
    field (orange arrows) displaces the coordinates $\rv_i$ of all
    particles $i$ by a vector $\eps(\rv_i)$; the specific particle $i$
    is highlighted in red and all particles are identical. A
    corresponding change in momenta, see
    \eqr{EQtransformationMomentum}, compensates the spatial distortion
    such that the differential phase space volume element (integration
    measure) remains unchanged.  (b)~Planar geometry of the confined
    Lennard-Jones fluid between two smooth parallel soft attractive
    Lennard-Jones walls; $\sigma$ is the particle size, $z$ measures
    the distance across the planar pore, and $L$ is the distance
    between the two walls.
\label{FIG1}}
\end{figure}

\subsection{Global shifting invariance}
\label{SECglobalShiftingTheory}

To develop the Noether invariance theory, we first consider a global
coordinate displacement $\rv_i \to \rv_i + \eps_0 \equiv \tilde\rv_i$,
where the shifting vector $\eps_0 = \rm const$ is independent of
position and acts on all particles in the same way
\cite{hermann2021noether}. The tilde indicates the new coordinates.
Expressing $H$ as well as the corresponding distribution function
$\Psi_\rmeq$ in the new coordinates makes averages become formally
dependent on the shifting parameter, i.e., $A(\eps_0)$. However, the
coordinate change can also be viewed as a mere re-parameterization of
the phase space integral which induces no change to its value such
that $A(\eps_0)=A$, with the right hand side denoting $\langle\hat
A\rangle$ in the original representation. Partially differentiating
both sides of the equation yields:
\begin{align}
  \frac{\partial A(\eps_0)}{\partial \eps_0}
  =\frac{\partial}{\partial\eps_0} \Tr \Psi_\rmeq(\eps_0) \hat A(\eps_0) 
  &= 0,
  \label{EQAparmatricDerivative}
\end{align}
where the second equality arises trivially from $\partial A/\partial
\eps_0=0$, as there is no dependence on $\eps_0$.  

Carrying out the derivative is straightforward upon using the
Boltzmann form of the equilibrium distribution function and noting
that the partition sum is independent of $\eps_0$. Explicitly we have
$\partial \Psi_\rmeq(\eps_0)/\partial \eps_0 = -\beta
\Psi_\rmeq(\eps_0) \partial H(\eps_0)/\partial \eps_0$.  Using the
product rule of differentiation and evaluating at vanishing
displacement $\eps_0=0$ then leads from \eqr{EQAparmatricDerivative}
to
\begin{align}
  -\beta\avg{\frac{\partial H(\eps_0)}{\partial \eps_0}
    \Big|_{\eps_0=0}\hat A} 
  +\avg{\frac{\partial \hat A(\eps_0)}{\partial \eps_0}
    \Big|_{\eps_0=0}} &= 0.
  \label{EQAparmatricDerivativeNextStep}
\end{align}
Observing that here $\partial/\partial\eps_0|_{\eps_0=0} \equiv \sum_i
\nabla_i$ allows to make the derivatives more explicit with $\nabla_i$
indicating the partial derivative with respect to $\rv_i$. In the
first term in \eqr{EQAparmatricDerivativeNextStep} we have $-\partial
H(\eps_0)/\partial \eps_0|_{\eps_0=0} = -\sum_i \nabla_i
V_\rmext(\rv_i) \equiv \hat \Fv_\rmext^0$, which defines the global
external force operator~$\hat \Fv_\rmext^0$. Here the global
interparticle force due to the mutual interactions between all
particles in the system vanishes, $\hat\Fv_\rmint^0=-\sum_i \nabla_i
u(\rv^N) \equiv 0$, as is due to Newton's third law or, analogously,
to the translational invariance of $u(\rv^N)$ against global
displacement~\cite{hermann2021noether}.

Re-ordering the two terms in \eqr{EQAparmatricDerivativeNextStep}
gives the following global hyperforce identity that holds for any
given observable $\hat A(\rv^N,\pv^N)$:
\begin{align}
  \beta \avg{\hat\Fv^0_\rmext \hat A} &= 
  -\avg{\sum_i\nabla_i \hat A}.
  \label{EQAepsGlobal}
\end{align}
Here $\hat A=\hat A(\rv^N,\pv^N)$ can feature additional parametric
dependence, such as on a generic position argument~$\rv$.  The sum
rule \eqref{EQAepsGlobal} relates the correlation of $\hat A$ with the
external force operator (left hand side) to the mean negative global
coordinate derivative of $\hat A$ (right hand side); here we use the
term correlation to imply the average of the product of two
observables.  As announced in the introduction, \eqr{EQAepsGlobal} is
similar to Hirschfelder's hypervirial theorem \cite{hirschfelder1960}
in the inclusion of the phase space function $\hat A(\rv^N, \pv^N)$
and our hyperforce terminology parallels his use of the term
hypervirial.

While Noether invariance enabled us to obtain the global hyperforce
identity \eqr{EQAepsGlobal} constructively, one can verify its
validity a posteriori by integration by parts in phase space on the
right hand side. The derivatives $\nabla_i$ then act on the
probability distribution and exploiting again the Boltzmann form leads
to $\hat A\sum_i \nabla_i \Psi_{\rm eq} = -\beta \Psi_{\rm eq} \hat A
\sum_i \nabla_i H = \Psi_{\rm eq} \beta\hat\Fv_\rmext^0 \hat A $,
which gives the left hand side of \eqr{EQAepsGlobal} upon building the
trace. Alternatively one can start with the Yvon theorem
\cite{hansen2013}, $\beta \langle \hat A \nabla_i H\rangle = \langle
\nabla_i \hat A \rangle$, which itself also follows from partial phase
space integration \cite{hansen2013}.  Summing the Yvon theorem over
all particles~$i$ and noting that $\sum_i \nabla_i u(\rv^N) \equiv 0$
gives \eqr{EQAepsGlobal}. A similar derivation can also be based on
the hypervirial theorem \cite{hirschfelder1960}.

\subsection{Local shifting invariance}
\label{SEClocalShiftingTheory}

Before presenting explicit applications of \eqr{EQAepsGlobal} to
specific forms of $\hat A$, we first generalize to the fully
position-resolved case. In generalization of the uniform coordinate
displacement of Sec.~\ref{SECglobalShiftingTheory}, we consider the
following local transformation on phase space, as parameterized by a
three-dimensional vector field $\eps(\rv)$ \cite{tschopp2022forceDFT,
  hermann2022quantum}:
\begin{align}
  \rv_i &\to \rv_i + \eps(\rv_i),
  \label{EQtransformationPosition}\\
  \pv_i &\to [\unity + \nabla_i\eps(\rv_i)]^{-1}\cdot \pv_i.
  \label{EQtransformationMomentum}
\end{align}
The gradient $\nabla_i \eps(\rv_i)$ is a $3\times 3$ matrix, $\unity$
denotes the $3\times 3$ identity matrix and the superscript $-1$
indicates matrix inversion.  Figure \ref{FIG1}(a) depicts an
illustration of the spatial transformation
\eqref{EQtransformationPosition}. The momentum transformation
\eqref{EQtransformationMomentum} has the following Taylor expansion to
lowest order in the displacement field: $\pv_i \to [\unity -
  \nabla_i\eps(\rv_i)]\cdot\pv_i$.

The joint transformation \eqref{EQtransformationPosition} and
\eqref{EQtransformationMomentum} is canonical
\cite{tschopp2022forceDFT, hermann2022quantum, goldstein2002} and it
hence preserves the phase space volume element, $d \rv_i d \pv_i = d
\tilde \rv_i d \tilde \pv_i$, where the tilde indicates the
transformed variables [right hand sides of
  Eqs.~\eqref{EQtransformationPosition} and
  \eqref{EQtransformationMomentum}]. The Hamiltonian also remains
unchanged (up to expressing the original via the new variables). Hence
the partition sum $\Xi$ is an invariant under the joint transformation
\eqref{EQtransformationPosition} and \eqref{EQtransformationMomentum}
\cite{tschopp2022forceDFT, hermann2022quantum,
  sammueller2023whatIsLiquid}.  Together with invariance of the
integration measure, the setup implies that any average $A=\langle
\hat A \rangle = \Tr \Psi_{\rm eq} \hat A$ is an invariant. This
property holds despite the explicit occurrence of the shifting field
$\eps(\rv)$ in the integrand, and hence $A[\eps]=A$, where the left
hand side carries the apparent dependence on the shifting field and
the right hand side is the average expressed in the original variables
where $\eps(\rv)$ is absent. We use standard notation to express
dependence on a function (so-called functional dependence) by
bracketed arguments.

From the local Noether invariance we can conclude from functionally
differentiating $A[\eps]=A$ with respect to the shifting field that
\begin{align}
  \frac{\delta A[\eps]}{\delta \eps(\rv)}   &= 0.
  \label{EQgenericAepsr}
\end{align}
The right hand side of \eqr{EQgenericAepsr} vanishes trivially due to
the average $A$ being independent of $\eps(\rv)$ in the original
representation and hence $\delta A/\delta \eps(\rv)=0$.  Carrying out
the functional derivative on the left hand side of
\eqr{EQgenericAepsr} requires to functionally differentiate the
equilibrium distribution, $\delta \Psi_{\rm eq}[\eps]/\delta \eps(\rv)
= -\beta \Psi_{\rm eq}[\eps] \delta H[\eps]/\delta \eps(\rv)$, as
follows from the chain rule. This allows to rewrite
\eqr{EQgenericAepsr} upon using the product rule and re-ordering the
resulting two terms as
\begin{align}
  -\beta\avg{\frac{\delta H[\eps]}{\delta \eps(\rv)}\Big|_{\eps=0}
    \hat A} =
  -\avg{\frac{\delta \hat A[\eps]}{\delta \eps(\rv)}\Big|_{\eps=0}},
  \label{EQgenericAepsr2}
\end{align}
where we have evaluated both sides at vanishing shifting field,
$\eps(\rv)=0$.

Differentiating the transformed Hamiltonian with respect to the
shifting field gives $ -\delta H[\eps]/\delta \eps(\rv)|_{\eps=0}=
\hat \Fv(\rv)$ \cite{tschopp2022forceDFT,hermann2022quantum}, where
the position-resolved total force operator comprises the following
three terms:
\begin{align}
  \hat \Fv(\rv) = \nabla\cdot \hat\taub(\rv) 
  +\hat \Fv_\rmint(\rv)
  -\hat\rho(\rv) \nabla V_{\rm ext}(\rv).
  \label{EQForceOperatorDefinition}
\end{align}
The right hand side of \eqr{EQForceOperatorDefinition} features the
one-body kinematic stress operator $\hat\taub(\rv) = -\sum_i
\delta(\rv-\rv_i)\pv_i \pv_i / m$ \cite{schmidt2022rmp}, the one-body
interparticle force density operator $\hat\Fv_\rmint(\rv)= -\sum_i
\delta(\rv-\rv_i) \nabla_i u(\rv^N)$ \cite{schmidt2022rmp}, the
standard form of the density operator $\hat \rho(\rv) = \sum_i
\delta(\rv-\rv_i)$ \cite{hansen2013, evans1979, schmidt2022rmp}, and
the external force field $-\nabla V_\rmext(\rv)$. (The force density
operator $\hat \Fv(\rv)$ defined in \eqref{EQForceOperatorDefinition}
also arises as the time derivative of the one-body current operator
\cite{schmidt2022rmp}.  In equilibrium the kinematic term in
\eqref{EQForceOperatorDefinition} reduces to a diffusive contribution:
$\nabla\cdot \langle \hat\taub(\rv)\rangle = -k_BT \nabla\rho(\rv)$;
we refer the Reader to Ref.\ \cite{schmidt2022rmp} for details.)
Equation \eqref{EQgenericAepsr2} can then be written upon carrying out
the functional derivatives and using \eqr{EQForceOperatorDefinition}
(on the left hand side) together with the chain rule (on the right
hand side), as the following hyperforce sum rule that holds for a
given form of $\hat A=\hat A(\rv^N,\pv^N)$:
\begin{align}
   \beta\avg{\hat\Fv(\rv) \hat A} &=
  -\avg{\sum_i\delta(\rv-\rv_i) \nabla_i \hat A}
 \notag\\  &\quad
  -\nabla \cdot \avg{\sum_i \delta(\rv-\rv_i)
  \frac{\partial \hat A}{\partial \pv_i}\pv_i},
\label{EQAepsLocal}
\end{align}
where $\hat A$ can again feature additional parametric dependencies,
such as on $\rv$.

For cases where the observable under consideration is independent of
the momenta, i.e., $\hat A \equiv \hat A(\rv^N)$, the second term on
the right hand side of \eqr{EQAepsLocal} vanishes and we obtain the
coordinate-based local hyperforce sum rule:
\begin{align}
  \beta\avg{\hat\Fv(\rv)\hat A(\rv^N)} &= 
  -\avg{\sum_i\delta(\rv-\rv_i)\nabla_i \hat A(\rv^N)}.
  \label{EQAepsCoordinatesOnly}
\end{align}

Equation \eqref{EQAepsCoordinatesOnly} can be viewed as a
generalization of the framework developed by Coles et
al.~\cite{coles2019}, where they consider observables of the specific
form $\hat A = \sum_i a_i \delta(\rv-\rv_i)$, where $a_i$ is a unique
property of particle $i$ only, such as e.g.\ its charge or, when
taking orientational degrees of freedom into account, its polarization
\cite{coles2019}.

As a consistency check, from integrating the locally resolved sum
rules \eqref{EQAepsLocal} and \eqref{EQAepsCoordinatesOnly} over
position $\rv$, i.e., applying $\int d\rv$ to both sides of these
equations, and observing that $\int d \rv \delta(\rv-\rv_i)=1$, one
retrieves the global Noether identity~\eqref{EQAepsGlobal}. Here the
global external force is the only remaining non-trivial global force
contribution, $\hat\Fv^0 \equiv \int d\rv\hat\Fv(\rv)=
\hat\Fv_\rmext^0$, as the global interparticle force vanishes due to
Newton's third law \cite{hermann2021noether} and there is also no
global diffusive effect due to vanishing boundary terms. Hence
\eqr{EQAepsGlobal} continues to hold upon replacing $\hat\Fv_\rmext^0$
by the global total force operator~$\hat \Fv^0$.  

As a further remark, by splitting off the kinetic term in
\eqr{EQAepsCoordinatesOnly}, its left hand side can be re-written as
$\beta \langle \hat\Fv(\rv) \hat A(\rv^N) \rangle= \beta \langle \hat
\Fv_U(\rv)\hat A(\rv^N)\rangle -\nabla \langle \hat\rho(\rv)\hat
A(\rv^N)\rangle$, with the potential force density operator being
given as the sum of interparticle and external contributions:
$\hat\Fv_U(\rv)=\hat \Fv_\rmint(\rv)-\hat\rho(\rv)\nabla
V_\rmext(\rv)$.

In summary, the local force decomposition into ideal, interparticle
and external contributions in \eqr{EQAepsCoordinatesOnly} allows to
obtain the following more explicit form, which holds, as we recall,
provided that $\hat A=\hat A(\rv^N)$ is independent of the momenta:
\begin{align}
  & \beta \avg{\hat\Fv_\rmint(\rv)\hat A}
  -\beta\avg{\hat\rho(\rv)\hat A}\nabla V_\rmext(\rv)\notag\\
  & \quad\qquad =\nabla\avg{\hat\rho(\rv)\hat A} 
 -\avg{\sum_i\delta(\rv-\rv_i)\nabla_i \hat A}.
  \label{EQAepsCoordinatesExplicitForm}
\end{align}
For a given explicit form of~$\hat A$, such as in the concrete
examples discussed below, the sum
rule~\eqref{EQAepsCoordinatesExplicitForm} connects the three
irreducible correlators $\langle \hat\Fv_\rmint(\rv)\hat A\rangle$,
$\langle \hat\rho(\rv)\hat A\rangle$, and $\langle
\sum_i\delta(\rv-\rv_i)\nabla_i\hat A\rangle$ in a formally exact and
nontrivial way with each other. Setting $\hat A=1$ recovers the YBG
equation (which we take to imply thermal averages being taken), as
then then last term in \eqr{EQAepsCoordinatesExplicitForm} vanishes
and the remaining terms constitute \eqr{EQybg}. Paralleling the naming
convention of the hypervirial theorem, which generalizes the standard
virial theorem to include a further observable,
\eqr{EQAepsCoordinatesExplicitForm} attains the status of a hyper-YBG
equation or hyperforce balance relationship. Concrete applications
thereof are shown below in Sec.~\ref{SECapplicationsLocal}.

The correlators on the left hand sides of the sum
rules~\eqref{EQAepsGlobal}, \eqref{EQAepsLocal}, and
\eqref{EQAepsCoordinatesOnly} also constitute covariances. We recall
that the covariance of two observables $\hat A$ and $\hat B$, as
defined via $\cov(\hat A, \hat B)= \langle \hat A \hat B \rangle -
\langle \hat A\rangle \langle \hat B \rangle$ measures the correlation
of the fluctuations of the two observables around their respective
mean. In the present case the mean force vanishes both globally,
$\langle \hat\Fv_\rmext^0\rangle =0$, and locally, $\langle \hat
\Fv(\rv)\rangle =0$. Hence we can formally subtract the vanishing
averages and re-express $\langle\hat\Fv_\rmext^0 \hat
A\rangle=\cov(\hat\Fv^0_\rmext, \hat A)$ as well as $\langle \hat
\Fv(\rv) \hat A \rangle = \cov(\hat\Fv(\rv), \hat A)$.  Besides the
conceptual difference between correlation and covariance, in practical
sampling schemes it can be beneficial to work with covariances rather
than correlations to reduce statistical noise, as we will demonstrate
further below.

That \eqr{EQAepsCoordinatesOnly} holds can again be verified a
posteriori by phase space coordinate integration by parts on the right
hand side. Due to the product rule two contributions result, one from
the Boltzmann factor: $\sum_i \delta(\rv-\rv_i)\nabla_i \Psi_\rmeq =
-\beta \Psi_\rmeq \sum_i \delta(\rv-\rv_i)\nabla_i H$, and one from
the Dirac distribution: $\Psi_\rmeq \sum_i \nabla_i \delta(\rv-\rv_i)
= - \Psi_\rmeq \nabla \hat\rho(\rv)$. Together with the factor $\hat
A$ their combination yields the left hand side of
\eqr{EQAepsCoordinatesOnly} upon identifying $\hat\Fv(\rv)$ via
\eqr{EQForceOperatorDefinition}. The more general \eqr{EQAepsLocal}
follows analogously upon integrating by parts also with respect to the
momenta.

\section{Results}
\label{SECapplications}
\subsection{Global shifting applications}
\label{SECapplicationsGlobal}

We turn to applications and hence consider concrete examples for the
general phase space function $\hat A$, which has remained so far
unspecified in the above generic hyperforce framework. We start with
investigating the global invariance \eqref{EQAepsGlobal}, which as we
demonstrate constitutes a powerful device both if $\hat A$ is a global
object or if it is locally resolved via dependence on a position. We
first consider the seemingly trivial case $\hat A=1$, for which of
course $\langle 1 \rangle=1$ due to the correct normalization of
$\Psi_\rmeq$. The right hand side of \eqr{EQAepsGlobal} vanishes and
we obtain $\Fv_\rmext^0 \equiv \langle \hat \Fv_\rmext^0 \rangle =0$,
i.e.\ the vanishing of the mean external force in equilibrium
\cite{hermann2021noether}. This is intuitively expected, as can be
seen by contradiction as follows. If the mean external force did not
vanish, then the system would start to move on average
\cite{hermann2022topicalReview} and hence it would not be in
equilibrium.

Adressing the global external force and hence setting $\hat
A=\hat\Fv_\rmext^0$ in \eqr{EQAepsGlobal} leads upon simplifying the
right hand side via $\langle \sum_i \nabla_i \hat \Fv_\rmext^0 \rangle
=- \langle \sum_i \nabla_i \nabla_i V_\rmext(\rv_i)\rangle$ to the
recently formulated global force-variance relationship $\beta\langle
\hat \Fv_\rmext^0 \hat \Fv_\rmext^0 \rangle = \int d\rv\rho(\rv)
\nabla \nabla V_\rmext(\rv)$ \cite{hermann2022variance}. Here the
auto-correlation of the global external force (left hand side) equals
up to a factor $\beta$ the mean external potential energy curvature
(right hand side).

By iteratively replacing $\hat A$ with the composite $\hat
\Fv_\rmext^0 \hat A$ in \eqr{EQAepsGlobal}, one can systematically
generate higher order sum rules, starting with $\beta \langle \hat
\Fv_\rmext^0 \hat \Fv_\rmext^0 \hat A \rangle = -\langle
\hat\Fv_\rmext^0 \sum_i \nabla_i \hat A\rangle - \langle \hat A \sum_i
\nabla_i\hat\Fv_\rmext^0 \rangle$, where the first term on the right
hand side allows repeated application of \eqr{EQAepsGlobal} and the
second term can be written via the external potential curvature. The
result is the following global second order hyperforce sum rule:
\begin{align}
  \beta^2 \langle \hat
  \Fv_\rmext^0 \hat \Fv_\rmext^0 \hat A \rangle 
  &= \langle \sum_{ij}
  \nabla_i \nabla_j\hat A\rangle + \langle \hat A \sum_i \nabla_i
  \nabla_i \beta V_\rmext(\rv_i) \rangle,
  \label{EQsumRuleGlobalSecondOrder}
\end{align}
where $\hat A=\hat A(\rv^N,\pv^N)$. The second term on the right hand
of \eqr{EQsumRuleGlobalSecondOrder} side can alternatively be written
as an integral over a correlation function as follows: $\beta\int d\rv
\langle \hat A\hat\rho(\rv)\rangle \nabla \nabla V_\rmext(\rv)$, where
$\langle \hat A \hat\rho(\rv)\rangle$ is the correlation of $\hat A$
and the local density operator. Alternatively to the present route via
\eqr{EQAepsGlobal}, the sum rule \eqref{EQsumRuleGlobalSecondOrder}
can equivalently be derived from second-order invariance of $\langle
\hat A \rangle$ against global shifting and hence calculating
$\partial^2 A(\eps_0)/\partial\eps_0 \partial\eps_0=0$.

Addressing locally resolved correlation functions on the basis of
\eqr{EQAepsGlobal} allows to access a higher degree of spatial
resolution. We first consider the case $\hat A=\hat\rho(\rv)$, which
leads upon re-writing the right hand side of \eqr{EQAepsGlobal} via
$-\sum_i \nabla_i\hat\rho(\rv)=- \sum_i \nabla_i \delta(\rv-\rv_i) =
\sum_i \nabla \delta(\rv-\rv_i)= \nabla \hat\rho(\rv)$ to the
following identity:
\begin{align}
  \beta \avg{ \hat \Fv_\rmext^0 \hat\rho(\rv) }  &= \nabla\rho(\rv),
  \label{EQrhoFextSumRule}
\end{align}
where we recall that on the right hand side $\rho(\rv)=\langle
\hat\rho(\rv)\rangle$ is the averaged density profile.  Hence building
the correlation of the density operator with the global external force
acts to spatially differentiate the density profile.  We recall that
the density gradient $\nabla\rho(\rv)$, as it occurs on the right hand
side of \eqr{EQrhoFextSumRule}, follows alternatively from the YBG
equation \eqref{EQybg}, which upon multiplication by $\beta$ attains
the form $\nabla\rho(\rv)=\beta \Fv_\rmint(\rv) +\beta
\Fv_\rmext(\rv)$, where the external force density is simply given as
$\Fv_\rmext(\rv)= -\rho(\rv)\nabla V_\rmext(\rv)$.

It is interesting to note that \eqr{EQrhoFextSumRule}, when written in
covariance form as $\beta \cov(\hat
\Fv_\rmext^0,\hat\rho(\rv))=\nabla\rho(\rv)$, mirrors closely the
structure of the thermodynamic identity $\beta
\cov(N,\hat\rho(\rv))=\partial\rho(\rv)/\partial \mu \equiv
\chi_\mu(\rv)$, with the local compressibility $\chi_\mu(\rv)$
\cite{evans2015jpcm,evans2019pnas,coe2022prl}. Here rather than the
spatial gradient, the thermodynamic parametric derivative with respect
to chemical potential occurs. Equation \eqref{EQrhoFextSumRule} can
also be viewed as the so-called inverse Lovett-Mou-Buff-Wertheim
(LMBW) relation $\beta\int d\rv' H_2(\rv,\rv') \nabla' V_\rmext(\rv')
= \nabla\rho(\rv)$ \cite{lovett1976,wertheim1976}, as is obtainable
from global translational invariance \cite{hermann2021noether}. Having
explicit results for the density covariance
$H_2(\rv,\rv')=\cov(\hat\rho(\rv),\hat\rho(\rv'))$ is however not
required in the much more straightforward
form~\eqref{EQrhoFextSumRule}. Furthermore, by summing only over
particle pairs with unequal indices we obtain the distinct identity
$\langle \hat\Fv_\rmext^0 \hat\rho(\rv)\rangle_{\rm
  dist}=\Fv_\rmint(\rv)$, which again relates seemingly very different
physical objects identically to each other.  The derivation is
straightforward by starting from \eqr{EQrhoFextSumRule}, subtracting
the self contribution which is the YBG equation \eqref{EQybg} in the
form $-\beta \rho(\rv)\nabla V_\rmext(\rv) =
-\beta\Fv_\rmint(\rv)+\nabla\rho(\rv)$, and dividing the result by
$\beta$.

To validate the Noether invariance theory and to investigate its
implications for the use in force sampling methods, we turn to
many-body simulations and consider the Lennrd-Jones (LJ) fluid as a
representative microscopic model.  The LJ pair potential $\phi(r)$
between two particles separated by a distance $r$ has the familiar
form $\phi(r)=4 \epsilon [ (\sigma/r)^{12}-(\sigma/r)^{6}]$ with the
energy scale $\epsilon$ and particle size $\sigma$ both being
constants.

As our above statistical mechanical derivations continue to hold
canonically, we sample both via adaptive Brownian dynamics (BD)
\cite{sammueller2021} with fixed number of particles, but also using
Monte Carlo simulations in the grand canonical ensemble. Spatial
inhomogeneity is induced by confining the system between two planar,
parallel LJ walls.  Each wall is represented by an external potential
contribution $V_{\rm wall}(z)$ that we choose to be identical to the
LJ interparticle potential $\phi(r)$, but instead of the radial
distance~$r$ evaluated as a function of the distance $z$ perpendicular
to the wall, $V_{\rm wall}(z)= 4 \epsilon
[(\sigma/z)^{12}-(\sigma/z)^{6}]$.  The joint potential of both walls
is then $V_\rmext(z) = V_{\rm wall}(z)+V_{\rm wall}(L-z)$, with $L$
indicating the separation distance between the two walls.  The
specific choice of 12-6-wall potential is made for convenience only
and it differs from the physically motivated 9-3-form (see
e.g.\ Ref.~\cite{evans2019pnas}).

The system is periodic in the two directions perpendicular to the
$z$-direction across the slit; a sketch is shown in
Fig.~\ref{FIG1}(b). The wall separation distance is chosen as
$L=10\sigma$, with $\sigma$ denoting the LJ particle size, and the
lateral box length is also set to $10\sigma$. The LJ potential is cut
and shifted with a cutoff distance of $2.5\sigma$. The reduced
temperature is $k_BT/\epsilon=2$ with $\epsilon$ denoting the LJ
energy scale. We use $N=200$ particles. Sampling is started after
$10^8$ time steps that are used for equilibration. The subsequent
sampling runlength is $3\times 10^8$ time steps which corresponds to
$~\sim 2000 \tau_B$, where $\tau_B = \gamma \sigma^2/\epsilon$ denotes
the Brownian timescale with~$\gamma$ being the friction constant.  All
results that we show for correlators are obtained from evaluation as
covariances. Subtracting the residual contribution from the product of
the two mean values helps to remove artifacts that occur due to finite
sampling.

\begin{figure}[!t]
\includegraphics[page=1,width=.95\columnwidth]{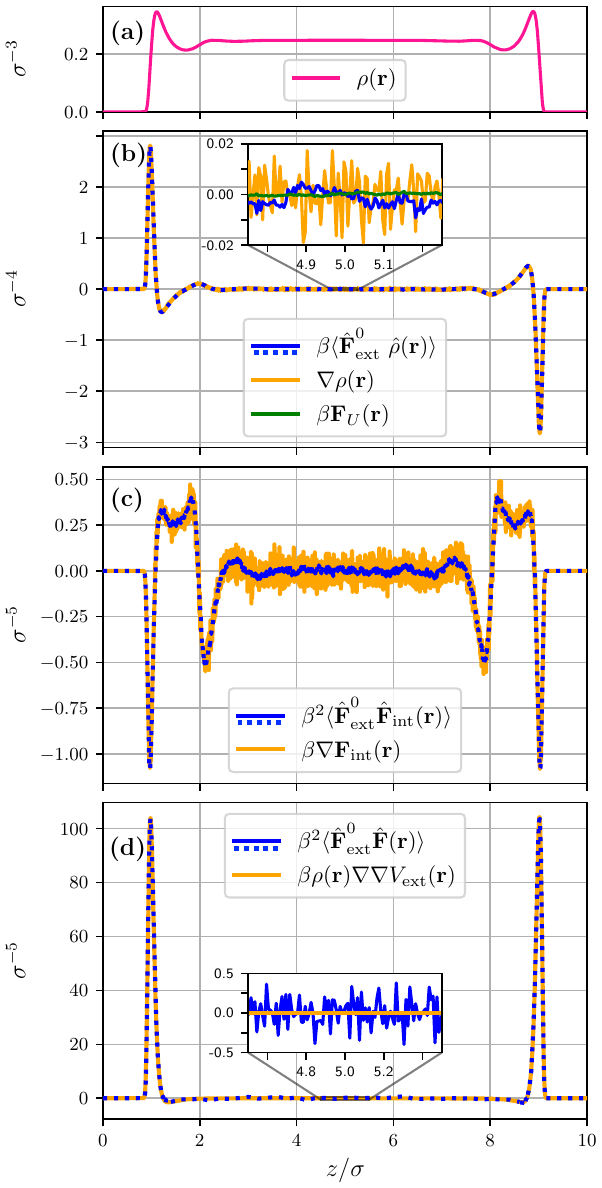}
\caption{ Illustration of the sum rules \eqref{EQrhoFextSumRule},
  \eqref{EQFintFextSumRule}, and \eqref{EQFextFtotSumRule}. The
  simulation results obtained from adaptive BD sampling of the LJ
  fluid confined between two parallel planar LJ walls. The profiles
  are shown as a function of scaled distance $z/\sigma$ across the
  planar slit.  (a) The density profile $\rho(\rv)$ of the confined
  system is shown as a reference.  (b) Comparison of the correlator
  $\beta \langle \hat\Fv_\rmext^0 \hat\rho(\rv)\rangle$ and the
  density gradient $\nabla\rho(\rv)$, see \eqr{EQrhoFextSumRule}; the
  zoomed inset demonstrates the respective noise levels and it also
  shows the scaled force density sum $\beta\Fv_U(\rv)=\beta
  \Fv_\rmint(\rv)+\beta\Fv_\rmext(\rv)$ which equals $\nabla\rho(\rv)$
  due to the local force balance.  (c) Comparison of $\beta^2 \langle
  \hat \Fv_\rmext^0 \hat\Fv_\rmint(\rv)\rangle$ and
  $\beta\nabla\Fv_\rmint(\rv)$, see \eqr{EQFintFextSumRule}. The
  former route carries less statisical noise and hence can serve a
  starting point for a force sampling scheme.  (d) Comparison of
  $\beta^2 \langle \hat \Fv_\rmext^0 \hat\Fv(\rv)\rangle$ and the
  local external potential curvature density
  $\beta\rho(\rv)\nabla\nabla V_\rmext(\rv)$, see
  \eqr{EQFextFtotSumRule}.}
\label{FIG2}
\end{figure}

The density profile of the confined LJ fluid, resolved as a function
of the scaled position $z/\sigma$ across the planar slit, is shown in
Fig.~\ref{FIG2}(a). The shape of the spatial density variation
features structured packing effects that appear adjacent to each wall
and that become damped towards the middle of the pore. Turning to the
density gradient, we present simulation results for both sides of
\eqr{EQrhoFextSumRule} in Fig.~\ref{FIG2}(b). 
Equation \eqref{EQrhoFextSumRule} in the specific planar geometry
reduces to $\beta \langle F_\rmext^{0,z}\sum_i\delta(z-z_i)/L^2\rangle
= \partial \rho(z)/\partial z$, where $\delta(z-z_i)$ is a
one-dimensional Dirac distribution, $z_i$ is the component of the
vector $\rv_i$ across the pore, $L^2$ is the lateral system area, and
the global external force has only a nonvanishing $z$-component given
by $F_\rmext^{0,z}=-\sum_i \partial V_\rmext(z_i)/\partial z_i$. The
density profile is sampled as $\rho(z) = \langle \sum_i
\delta(z-z_i)/L^2\rangle$,
The comparison of the a priori very different data sets shown in
Fig.~\ref{FIG2}(b) indicates excellent agreement. That the correlation
of the density operator with the global external force operator indeed
gives the gradient of the density profile, cf.~\eqr{EQrhoFextSumRule},
is surely not only at first glance very counter-intuitive.

We next consider the one-body interparticle force density operator
$\hat A = \hat\Fv_\rmint(\rv)$, for which \eqr{EQAepsGlobal} yields
\begin{align}
  \beta \avg{\hat\Fv_\rmext^0 \hat\Fv_\rmint(\rv)}
  &= \nabla \Fv_\rmint(\rv).
  \label{EQFintFextSumRule}
\end{align}
Quite remarkably, \eqr{EQFintFextSumRule} gives access to the gradient
of the internal force density (right hand side) via sampling the
correlation of the local internal force density with the global
external force (left hand side). This relationship could be used in a
force sampling scheme \cite{delasheras2018forceSampling, borgis2013,
  purohit2019, rotenberg2020, coles2019, coles2021,
  coles2023revelsMD}, where one obtains data for the force
correlations and via spatial integration obtains the interparticle
force density, which we demonstrate below in
Sec.~\ref{SECforceSampling}. We first present simulation results to
illustrate the validity of \eqr{EQFintFextSumRule} in
Fig.~\ref{FIG2}(c). Remarkably, the results for $\beta\langle
\hat\Fv_\rmext^0 \hat\Fv_\rmint(\rv)\rangle $ carry much less
statistical noise, as the need for building the numerical derivative
$\nabla \Fv_\rmint(\rv)$ is avoided.  However, this is no panacea, as
accurately sampling the correlation of the interparticle force density
with the global external force also poses challenges to overall
equilibration of the system.

Addressing the total force density operator $\hat A =\hat\Fv(\rv)$
requires to complement the above considered interparticle force
density~$\hat\Fv_\rmint(\rv)$ with the ideal and external
contributions. These two latter terms constitute mere variants of the
density operator identity \eqref{EQrhoFextSumRule}.  First there is
the external force density operator $\hat A = \hat \Fv_\rmext(\rv) =
-\hat\rho(\rv) \nabla V_\rmext(\rv)$ which yields $\beta\langle\hat
\Fv^0_\rmext \hat \Fv_\rmext(\rv)\rangle = -[\nabla
  V_\rmext(\rv)]\nabla\rho(\rv) $, as $\nabla V_\rmext(\rv)$ can be
taken out of the phase space average on the left hand side of
\eqr{EQrhoFextSumRule}. Secondly, the diffusive force density, $\hat A
= -k_BT \nabla\hat\rho(\rv)$, leads trivially to the gradient of
\eqr{EQrhoFextSumRule}.

Collecting all three terms (ideal, interparticle, and external) allows
to obtain for the choice $\hat A = \hat \Fv(\rv)$ a mixed global-local
Noether identity:
\begin{align}
  \beta\avg{\hat\Fv_\rmext^0 \hat \Fv(\rv)}
  &= \rho(\rv) \nabla\nabla V_\rmext(\rv).
  \label{EQFextFtotSumRule}
\end{align}
We present simulation results that validate the sum rule
\eqref{EQFextFtotSumRule} in Fig.~\ref{FIG2}(d). We have checked that
via spatial integration these results also validate the global
variance identity $ \beta\langle \hat\Fv_\rmext^0
\hat\Fv_\rmext^0\rangle = \int d\rv\rho(\rv) \nabla\nabla
V_\rmext(\rv)$ \cite{hermann2022variance}, which follows from
\eqr{EQsumRuleGlobalSecondOrder} with operator $\hat A=1$.  For the
presently considered system the global force-force correlation on the
left hand side is only marginally ($0.4\%$) smaller than the global
mean potential curvature on the right hand side.

\begin{figure}[!t]
\includegraphics[page=1,width=.99\columnwidth]{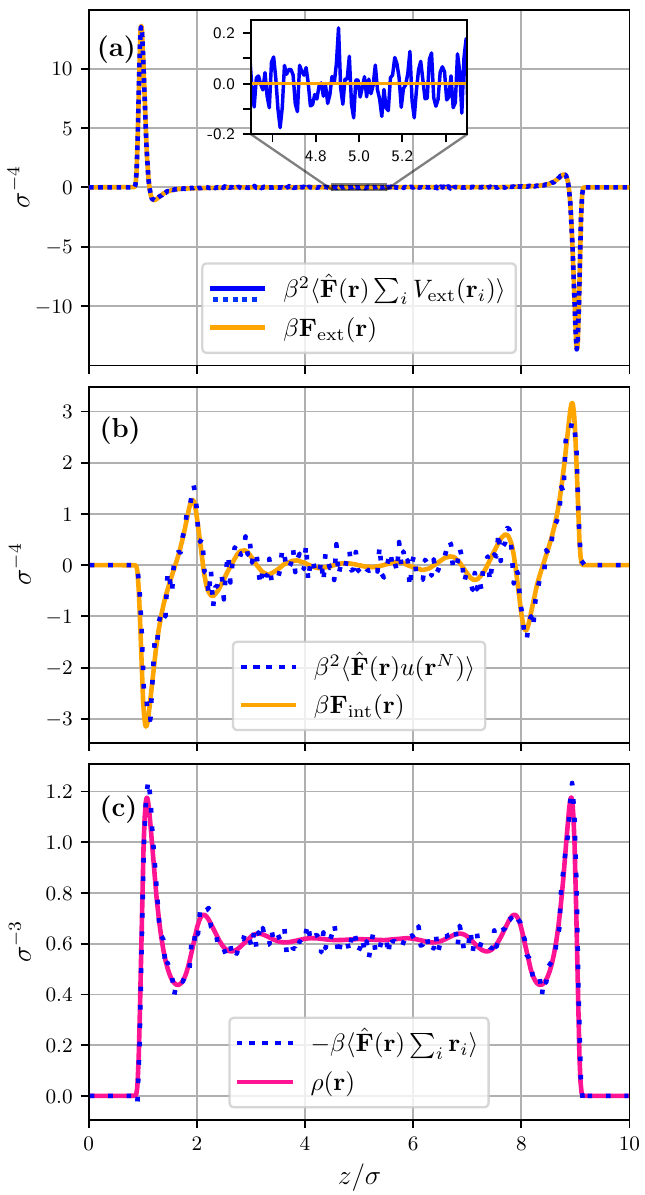}
\caption{Demonstration of Noether sum rules
  \eqref{EQexternalEnergySumRule}, \eqref{EQinternalEnergySumRule},
  and \eqref{EQcenterOfMassSumRule}. These identites are respectively
  based on the global external and interparticle energies, and on the
  center of mass. Shown are results from adaptive BD simulations for
  the LJ fluid between parallel LJ walls as a function of the scaled
  distance $z/\sigma$.  (a) Comparison of $\beta^2\langle \hat
  \Fv(\rv) \sum_i V_\rmext(\rv_i)\rangle$ and $\beta
  \Fv_\rmext(\rv)=-\beta \rho(\rv)\nabla V_\rmext(\rv)$, see
  \eqr{EQexternalEnergySumRule}.  (b) Comparison of
  $\beta^2\langle\hat\Fv(\rv) u(\rv^N)\rangle$ and $\beta
  \Fv_\rmint(\rv)$, see \eqr{EQinternalEnergySumRule}.  (c) Comparison
  of $-\beta\langle \hat \Fv(\rv) \sum_i \rv_i\rangle$ and
  $\rho(\rv)$, see \eqr{EQcenterOfMassSumRule}.
  \label{FIG3}}
\end{figure}

\subsection{Local shifting applications}
\label{SECapplicationsLocal}

We turn to application of the locally resolved hyperforce sum
rules~\eqref{EQAepsLocal} and \eqref{EQAepsCoordinatesOnly}. We recall
from Sec.~\ref{SEClocalShiftingTheory} that the seemingly trivial case
$\hat A=1$ reduces the local hyperforce identity
\eqref{EQAepsCoordinatesOnly} to the locally resolved force balance
relationship $\Fv(\rv)=\langle \hat\Fv(\rv) \rangle = 0$.  The
definition of the total force operator
\eqref{EQForceOperatorDefinition} and carrying out the average gives
the more explicit form $-k_BT \nabla \rho(\rv) + \Fv_\rmint(\rv) -
\rho(\rv)\nabla V_\rmext(\rv)=0$, i.e., the first member \eqref{EQybg}
of the Yvon-Born-Green hierarchy \cite{hansen2013}.  Typically this
identity is derived from multiplying the equilibrium probability
distribution function $\Psi_\rmeq$ by the gradient of the interaction
potential and integrating over the degrees of freedom of $N-1$
particles. We emphasize that arguably the simplest possible
application of \eqr{EQAepsCoordinatesOnly} yields such a central
result of liquid state theory with very little effort. From the
Noetherian point of view the result was also obtained from applying
the locally-resolved transformation \eqref{EQtransformationPosition}
and \eqref{EQtransformationMomentum} to the free energy
\cite{tschopp2022forceDFT, hermann2022quantum}. At the heart of these
Noetherian derivations lies the invariance of the Hamiltonian, of the
phase space integration measure, and hence of the partition sum.

We next consider setting $\hat A = \hat\Fv_\rmext^0$ in
\eqr{EQAepsLocal}. This constitutes a valuable consistency check with
the above global shifting of $\hat\Fv(\rv)$ that led to
\eqr{EQFextFtotSumRule} and which we can identically reproduce here.
Arguably even more fundamentally the sum rule
\eqref{EQFextFtotSumRule} can be obtained by considering mixed
local-global shifting invariance at second order, i.e., building the
mixed derivative $\partial (\delta
\Omega/\delta\eps(\rv))/\partial\eps_0=0$, where the grand potential
$\Omega=-k_BT\ln\Xi$ is subject to the combined displacement $\rv_i
\to \rv_i + \eps_0 + \eps(\rv_i)$.  Furthermore increasing the spatial
resolution and hence selecting $\hat A = \hat \Fv(\rv)$ in
\eqr{EQAepsLocal} yields the recent Noether-constrained two-body
force-correlation theory, which is discussed in detail in
Refs.~\cite{sammueller2023whatIsLiquid}.  The theory that is presented
therein can hence be viewed as the special case of pure
force-dependence within the hyperforce framework.

Reverting back to developing the general theory, here we generalize to
higher orders by iteratively replacing $\hat A(\rv^N,\pv^N)$ by $\hat
\Fv(\rv') \hat A(\rv^N,\pv^N)$ in \eqr{EQgenericAepsr2}. This leads to
the following second order hyperforce sum rule:
\begin{align}
  \beta^2\langle
  \hat\Fv(\rv)\hat\Fv(\rv')\hat A\rangle
  &= \beta \avg{\hat A 
  \frac{\delta^2 H[\eps]}{\delta\eps(\rv)\delta\eps(\rv')}}
  +\avg{
    \frac{\delta^2 \hat A[\eps]}{\delta\eps(\rv)\delta\eps(\rv')}
  },
  \label{EQsecondOrderHyperForce}
\end{align}
where evaluation of the right hand side at $\eps(\rv)=0$ is suppressed
in the notation and $\delta^2 H[\eps]/\delta \eps(\rv)\delta
\eps(\rv')$ is discussed in Ref.~\cite{sammueller2023whatIsLiquid}.
In case of no dependence of $\hat A$ on momenta, i.e.\ $\hat A=\hat
A(\rv^N)$, the second term on the right hand side of
\eqr{EQsecondOrderHyperForce} can be made more explicit as
\begin{align}
  \avg{\frac{\delta^2 \hat A([\eps],\rv^N)}{\delta
    \eps(\rv)\delta\eps(\rv')}}
  &=\avg{
  \sum_{ij}\delta(\rv-\rv_i)\delta(\rv'-\rv_j)\nabla_i \nabla_j \hat
  A(\rv^N)}.
\end{align}

\begin{figure*}[!t]
  \includegraphics[page=1,width=1.6\columnwidth]{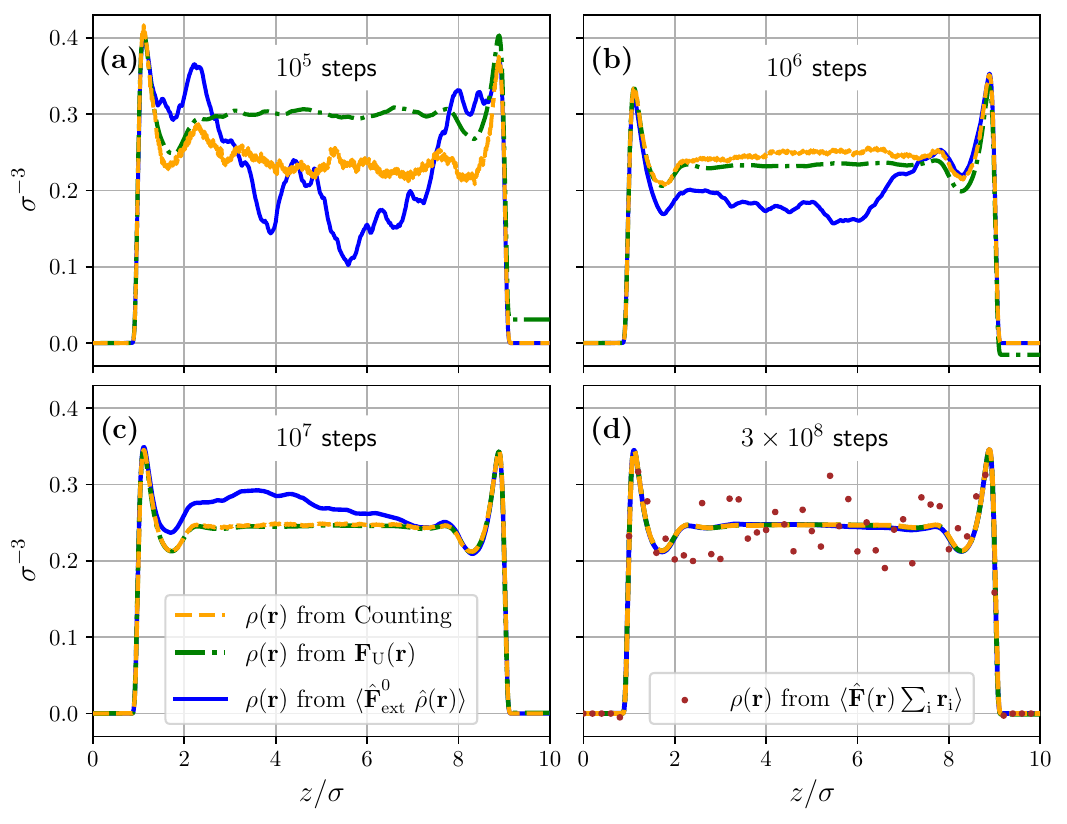}
    \caption{Comparison of standard counting against force sampling.
      We show results from four different routes towards the density
      profile $\rho(\rv)$ as obtained after (a) $10^5$, (b) $10^6$,
      (c) $10^7$, (d) $3\times 10^8$ simulation steps.  Shown is data
      from the standard counting method according to
      \eqr{EQforceSamplingDensityProfilePlanar1} (orange lines), from
      force sampling $\Fv_U(\rv)$ according to
      \eqr{EQforceSamplingDensityProfilePlanar2} (green dash-dotted
      lines), from global external force correlation sampling of
      $\langle \hat \Fv_\rmext^0 \hat\rho(\rv)\rangle$ according to
      \eqr{EQforceSamplingDensityProfilePlanar3} (blue solid lines),
      and from center-of-mass correlation sampling of
      $\langle\hat\Fv(\rv)\sum_i\rv_i \rangle$ according to
      \eqr{EQforceSamplingDensityProfilePlanar4} (red
      symbols). Results from the latter route are only displayed from
      the longest run [panel (d)] and they still display considerable
      scatter, whereas the results from the remaining three methods
      already agree very satisfactorily.}
  \label{FIG4}
\end{figure*}

\begin{figure*}[!t]
\includegraphics[page=1,width=1.6\columnwidth]{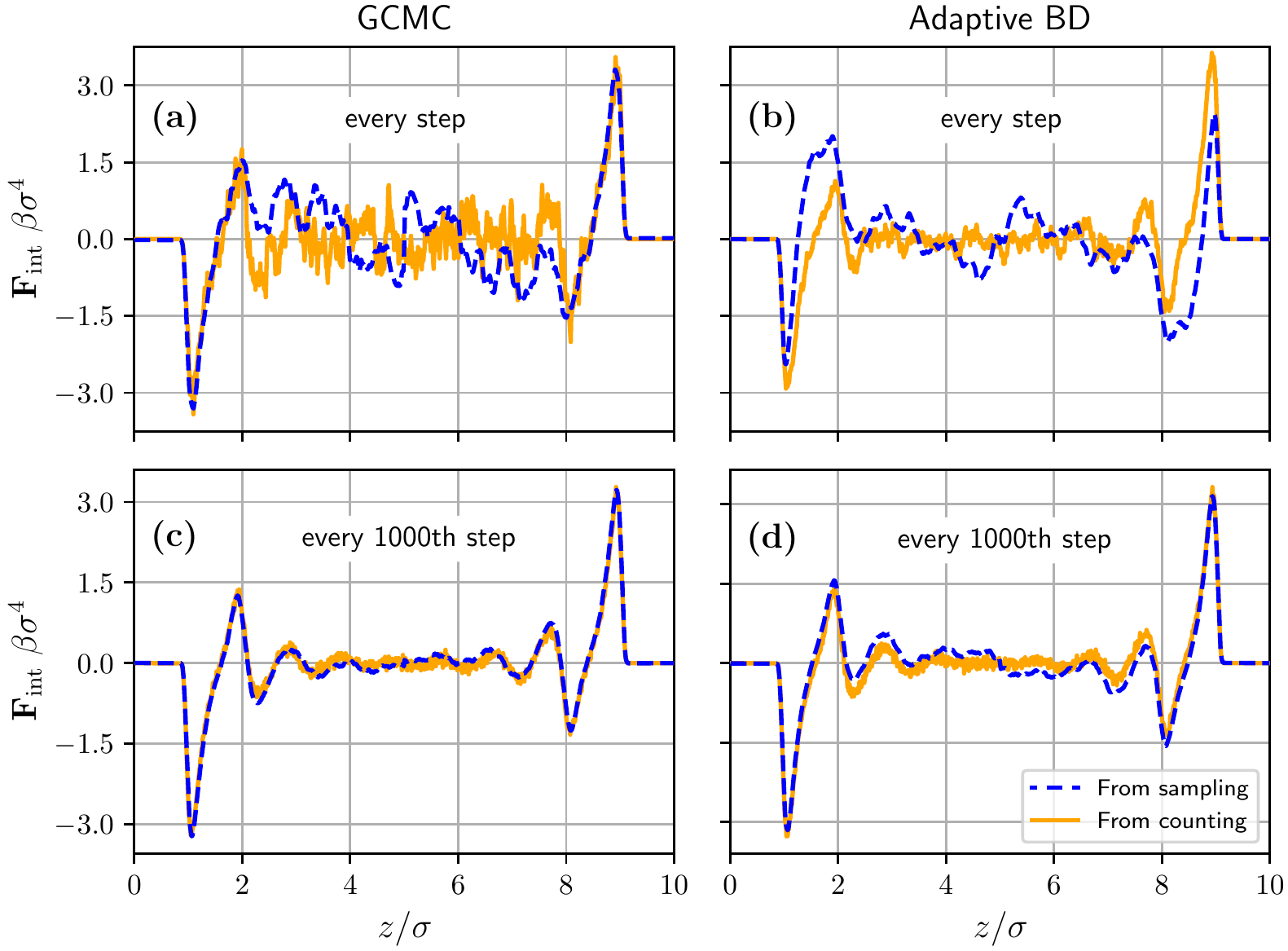}
  \caption{Comparison of different statistical estimators for the
    interparticle one-body force density profile according to
    \eqr{EQFintFextSumRule}. The results are obtained from the
    standard counting histogram method, $\Fv_\rmint(\rv)= \langle
    \hat\Fv_\rmint(\rv)\rangle$ (orangle lines), and from hyperforce
    sampling and spatial integration of $\nabla
    \Fv_\rmint(\rv)=\beta\langle
    \hat\Fv_\rmext^0\hat\Fv_\rmint(\rv)\rangle$ (blue lines) with data
    for the right hand side forming the basis. These methods are
    explicitly spelled out in Eqs.~\eqref{EQhyperforceCounting} and
    \eqref{EQhyperforceSampling}.  Results are shown from grand
    canonical Monte Carlo simulations [panels (a) and (c)] and from
    adaptive BD simulations [(b) and (d)].  The results stem from
    sampling $10^5$ strongly correlated microstates at every
    simulation step [(a) and (b)] compared to better decorrelated
    configurations obtained from also $10^5$ configurations, with the
    samples being taken only every 1000th simulation step [(c) and
      (d)].  The results from sampling are symmetrized with respect to
    mirroring at the center of the pore, i.e., via building the
    arithmetic mean $[\Fv_\rmint(z) - \Fv_\rmint(L-z)]/2$, as is
    common practice in force sampling schemes.}
  \label{FIG5}
\end{figure*}

We proceed beyond forces by turning to energies with the aim of
exploiting their thermal Noether invariance against shifting.  We
consider both the global external potential energy $\hat A = \sum_i
V_\rmext(\rv_i)$ as well as the global interparticle energy $\hat
A=u(\rv^N)$.  Applying \eqr{EQAepsCoordinatesOnly} yields in these two
cases respectively the following sum rules:
\begin{align}
  \beta \avg{\hat\Fv(\rv) \sum_i V_\rmext(\rv_i)} &=
  -\rho(\rv) \nabla V_\rmext(\rv),
  \label{EQexternalEnergySumRule}  \\
  \beta \avg{\hat \Fv(\rv) u(\rv^N)} &= \Fv_\rmint(\rv).
  \label{EQinternalEnergySumRule}
\end{align}
Simulation results that demonstrate the validity of
Eqs.~\eqref{EQexternalEnergySumRule} and
\eqref{EQinternalEnergySumRule} are shown in Fig.~\ref{FIG3}(a) and
(b), respectively.  Here we have increased the overall density by
reducing the lateral box size to $5\sigma$ and we sample $N=128$
particles over time periods of $2000 \tau$ [data shown in
  Figs.~\ref{FIG3}(a) and (b)] and of $8000 \tau$ [data shown in
  Figs.~\ref{FIG3}(c)].

The potential energy identities \eqref{EQexternalEnergySumRule} and
\eqref{EQinternalEnergySumRule} can be supplemented by considering the
kinetic energy, $\hat A=\sum_i \pv_i^2/(2m)$, which leads upon using
\eqr{EQAepsLocal} to the identity $\beta \langle \hat \Fv(\rv) \sum_i
\pv_i^2/(2m)\rangle = -k_BT \nabla\rho(\rv)$.  Treating then the
entire Hamiltonian, $\hat A=H$, follows from adding up all three
energy contributions. The result is the compact identity:
$\beta\langle \hat\Fv(\rv)H\rangle=\Fv(\rv)=0$.  This possibly
unexpected behaviour also holds for the global entropy. We choose the
entropy operator $\hat A = \hat S \equiv -k_B \ln \Psi_\rmeq$ and
obtain from \eqr{EQAepsLocal} $k_B^{-1}\langle \hat \Fv(\rv)\hat
S\rangle=\Fv(\rv)=0$, i.e., the correlation (as well as the
covariance) of the entropy operator with the local force density
vanishes. This behaviour is very different from the nontrivial
fluctuation profile that is obtained from the covariance of the
density operator with the global entropy \cite{eckert2020,
  eckert2023fluctuation}.

As a final case we consider the center of mass $\sum_i \rv_i/N$ as a
purely mechanical entity. We multiply by $N$, such that $\hat
A=\sum_i\rv_i$ and obtain from \eqr{EQAepsCoordinatesOnly} upon
multiplication by $-1$ the result
\begin{align}
  -\beta \avg{\hat\Fv(\rv)\sum_i \rv_i} &= 
  \rho(\rv) \unity.
  \label{EQcenterOfMassSumRule}
\end{align}

Integrating \eqr{EQcenterOfMassSumRule} over position yields a simple
relationship of the correlator of the global external force and the
center of mass: $\langle \hat\Fv_\rmext^0\sum_i\rv_i\rangle/\bar N =
-k_BT \unity$, where the mean number of particles is $\bar N = \int
d\rv \rho(\rv)\equiv \langle N \rangle$.  Except for an additional sum
over all particles, this global relationship is akin to the
equipartition theorem.  We present simulation results for both sides
of the locally resolved \eqr{EQcenterOfMassSumRule} in
Fig.~\ref{FIG3}(c). The accurate agreement of the respective profiles
confirms that the identity \eqref{EQcenterOfMassSumRule} indeed offers
a rather unusual route to gain access to the density profile.

\subsection{Hyperforce sampling and equilibration testing}
\label{SECforceSampling}

Besides the unexpected insights into the general correlation structure
of equilibrium many-body systems that the thermal Noether invariance
delivers, our results are useful for the careful assessment and
construction of computer sampling schemes. Force sampling
\cite{rotenberg2020, borgis2013, delasheras2018forceSampling,
  purohit2019, coles2019, coles2021, coles2023revelsMD} in perhaps its
most intuitive form \cite{delasheras2018forceSampling} rests on
spatial integration of the YBG equation \eqref{EQybg} such that the
density profile is obtained via $\rho(\rv)=\rho_0 + \beta
\nabla^{-1}\cdot[\Fv_\rmint(\rv)-\rho(\rv)\nabla V_\rmext(\rv)]$,
where $\rho_0=\rm const$ is an integration constant and $\nabla^{-1}$
is an inverse $\nabla$ operator. The data input on the right hand side
is obtained via sampling $\Fv_\rmint(\rv)=\langle
\hat\Fv_\rmint(\rv)\rangle$ and either
$\rho(\rv)=\langle\hat\rho(\rv)\rangle$ or $\Fv_\rmext(\rv)=-\langle
\sum_i \delta(\rv-\rv_i)\nabla_i V_\rmext(\rv_i) \rangle$. The
averages denote those that are being carried out in the simulation. In
the present planar geometry $\nabla^{-1}$ reduces to carrying out a
simple position integral, which we make explicit below.

Summarizing, we can compare the results from four different routes: i)
counting of particle occurrences in a position-resolved histogram,
which constitutes the standard method; ii) force sampling
\cite{delasheras2018forceSampling} according to \eqr{EQybg}, iii)
hyperforce sampling according to the global external force correlation
in \eqr{EQrhoFextSumRule} with spatial integration post processing,
and iv) center-of-mass-based hyperforce sampling according to
\eqr{EQcenterOfMassSumRule}. These routes are respectively given by
the following explicit expressions:
\begin{align}
  \rho(z) &= \langle \hat\rho(z) \rangle,
  \label{EQforceSamplingDensityProfilePlanar1}  \\
  \rho(z) &= \rho_0 + \beta \int_0^z dz' 
      \avg{\hat F_U(z')},
      \label{EQforceSamplingDensityProfilePlanar2}\\
  \rho(z) &= \rho_0 + \beta \int_0^z dz' 
  \avg{ \hat F^0_{\rmext\commaz}\hat\rho(z') },
      \label{EQforceSamplingDensityProfilePlanar3}\\
  \rho(z) &= -\beta \avg{\hat F_{\commaz}(z) \sum_i z_i}.
      \label{EQforceSamplingDensityProfilePlanar4}
\end{align}
Here we choose the integration constant as $\rho_0=\rho(0)=0$ due to
the divergent wall potential at $z=0$.  The averages on the above
right hand sides denote the actual simulation data, all vectors have
been projected onto the $z$-direction across the pore, and $z_i$
denotes the $z$-component of the particle position~$\rv_i$.  In more
detail, the operators on the right hand sides of
Eqs.~\eqref{EQforceSamplingDensityProfilePlanar1} and
\eqref{EQforceSamplingDensityProfilePlanar2} are explicitly given as
$\hat\rho(z)=\sum_i \delta(z-z_i)/L^2$ and $\hat F_U(z)=\sum_i
f_{i,z}^\rmint \delta(z-z_i)/L^2 - \hat\rho(z) \partial
V_\rmext(z)/\partial z$, where we recall that $L^2$ is the lateral
system size and the $z$-component of the interparticle force on
particle $i$ is $f_{i,z}^\rmint = -\partial u(\rv^N)/\partial z_i$.
Furthermore the operators on the right hand sides of
Eqs.~\eqref{EQforceSamplingDensityProfilePlanar3} and
\eqref{EQforceSamplingDensityProfilePlanar4} are $\hat F_\rmext^0 =
-\sum_i \partial V_\rmext(z_i)/\partial z_i$, and $\beta \hat F(z)=
\beta \hat F_U(z)-\partial\hat\rho(z)/\partial z$.

Results for the density profile from the four routes
\eqref{EQforceSamplingDensityProfilePlanar1}-\eqref{EQforceSamplingDensityProfilePlanar4}
are shown in Fig.~\ref{FIG4}. The simulation parameters are identical
as before [Fig.~\ref{FIG2}]. We display the four different statistical
estimators for the density profile, as obtained after increasing
runlength of (a) $10^5$, (b) $10^6$, (c) $10^7$, and (d) $3\times
10^8$ simulation steps. We recall that as demonstrated above both in
the numerical examples as well as in the formal statistical mechanical
derivations, the results from all routes are formally identical. In
practice, pronounced differences can be observed and these are due to
the simulation averages being mere approximations for the true
statistical mechanical equilibrium.

For example the routes \eqref{EQforceSamplingDensityProfilePlanar2}
and \eqref{EQforceSamplingDensityProfilePlanar3} yield less
statistical noise due to the spatial integration, but they however can
instead accumulate systematic deviations.  The expexted differences
between the four methods are also consistently demonstrated by the
fact that the results from the different routes mutually agree better
for increasing runlengths. Nevertheless, in particular the routes
\eqref{EQforceSamplingDensityProfilePlanar3} and
\eqref{EQforceSamplingDensityProfilePlanar4} that involve global
quantities are very sensitive to choice of runlength and they can
hence serve as indicators of the overall quality of the sampling
routine, even when quantities beyond the density profile are the very
aim of the simulation. Having such tools for quality control can be
particularly useful when investigating capillary and wetting phenomena
\cite{upton1998, evans1990, henderson1984, henderson1985,
  triezenberg1972} where surface phase transitions pose significant
challenges for reliable prediction.

The Noetherian hyperforce framework allows us to easily go beyond the
density profile and we wish to address the interparticle force density
as a target, rather than the mere source that it played in
contributing to \eqr{EQforceSamplingDensityProfilePlanar2} above for
the force sampling. As a demonstration we use and contrast different
estimators for the interparticle force density profile
$\Fv_\rmint(\rv)$. The traditional counting method of filling a
position-resolved histogram forms the baseline and
\eqr{EQFintFextSumRule} provides an alternative. These methods are
respectively given by
\begin{align}
  F_{\rmint\commaz}(z) &= \langle \hat F_\rmint(z) \rangle,
  \label{EQhyperforceCounting}\\
  F_{\rmint\commaz}(z) &= 
  \beta\int_0^z dz'
  \avg{\hat F_{\rmext\commaz}^0 \hat F_{\rmint\commaz}(z')}.
  \label{EQhyperforceSampling}
\end{align}

Figure~\ref{FIG5} presents both canonical averages obtained via
adaptive BD \cite{sammueller2021} as well as grand canonical Monte
Carlo data. The chemical potential is chosen as $\mu/\epsilon=1$ and
the resulting average number of particles is $\langle
N\rangle=136.5$. In the corresponding adaptive BD simulation runs we
have set $N=136$, which remains sharply fixed in the course of time.
The agreement between both sets of results confirms the expectation of
independence of the sum rule validity on the choice of ensemble.  This
is based on the fact that the theoretical derivations continue to hold
with fixed $N$, as we have also explicitly verified. Hence the
mechanical effects that the Noether invariance against spatial
displacement captures are oblivious to the presence of global particle
number fluctuations. We recall that the later are precisely captured
and quantified by the local compressibility
\cite{evans2015jpcm,evans2019pnas,coe2022prl}.

The comparison of lower and higher quality statistical data, as
obtained from sampling every step (top row) or only every 1000th step
(bottom row) demonstrates that the force correlation method is a
sensitive measure of the degree of sampling quality.

\section{Conclusions}
\label{SECconclusions}

\begin{figure}[!t]
\includegraphics[width=0.99\columnwidth, page=1]{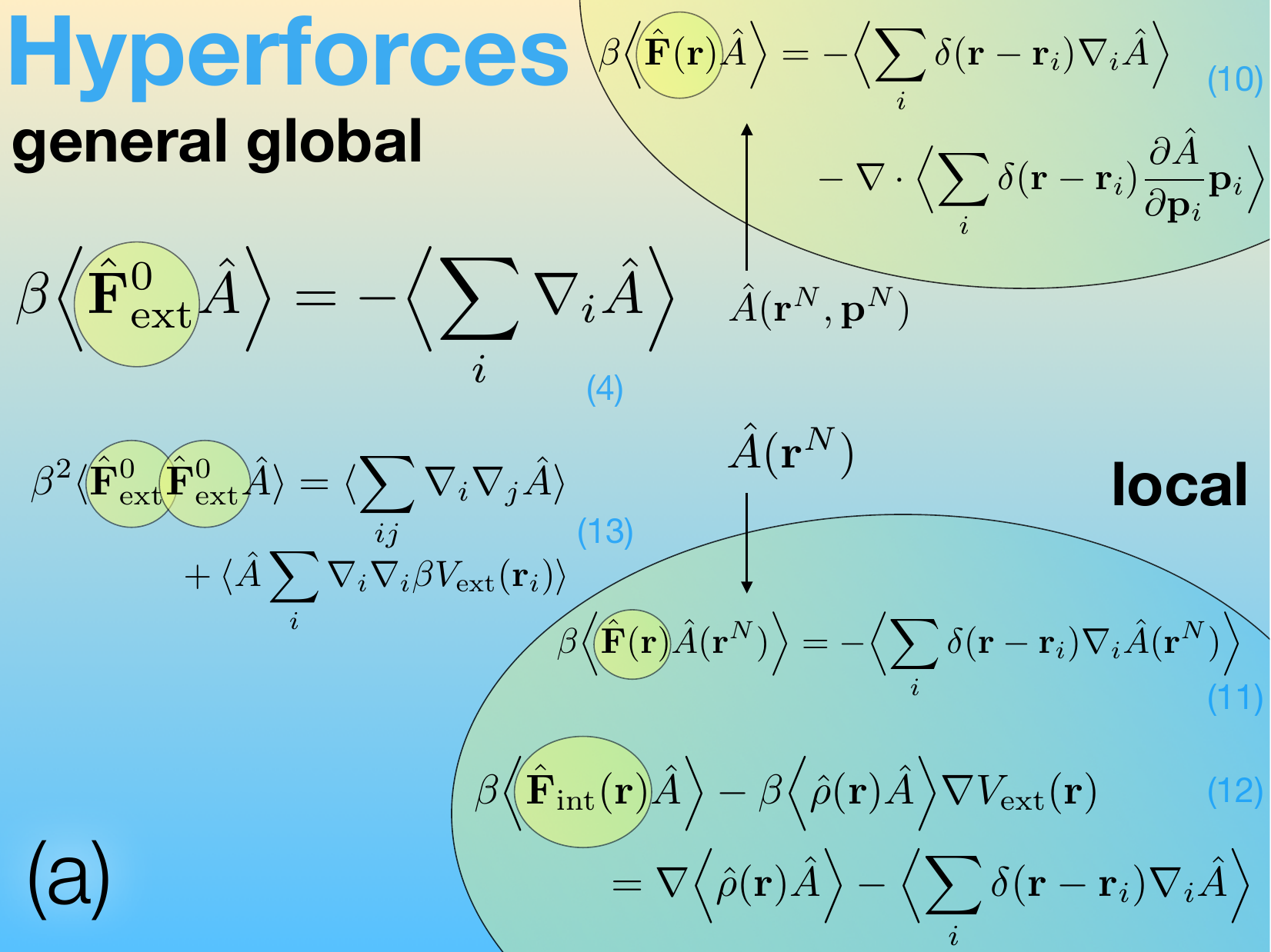}
\includegraphics[width=0.99\columnwidth, page=1]{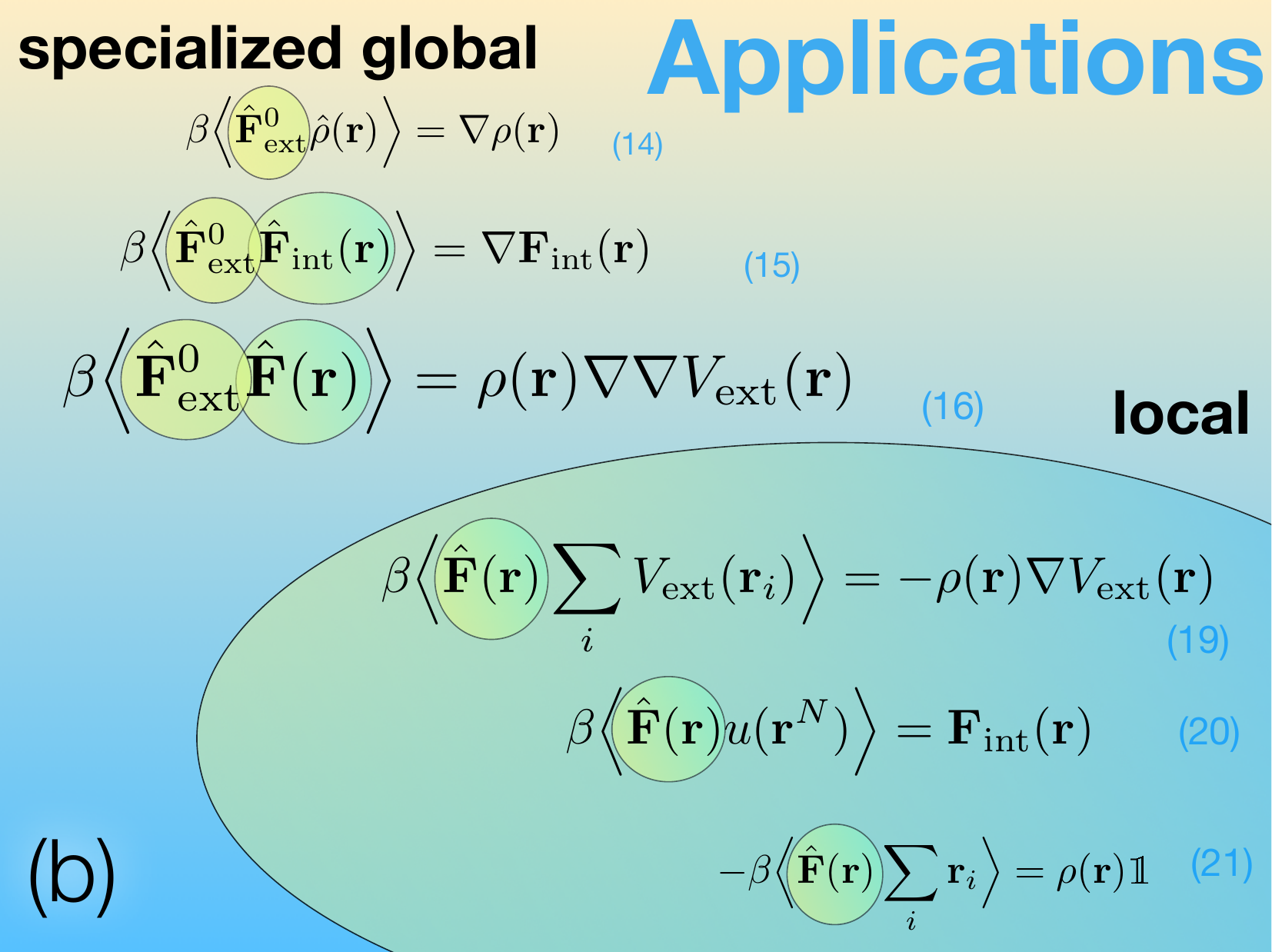}
  \caption{Overview of theoretical results. (a) Hyperforce sum rules
    that relate a given observable with the forces that act in the
    system. The global sum rules \eqref{EQAepsGlobal} and
    \eqref{EQsumRuleGlobalSecondOrder} couple the global external
    force operator $\hat\Fv_\rmext^0$ at first and second order with
    an observable $\hat A(\rv^N, \pv^N)$.
    The local sum rules \eqref{EQAepsLocal},
    \eqref{EQAepsCoordinatesOnly}, and
    \eqref{EQAepsCoordinatesExplicitForm} contain the localized force
    density operator $\hat\Fv(\rv)$ and its interparticle contribution
    $\hat\Fv_\rmint(\rv)$ and they apply for full phase
    space-dependence $\hat A(\rv^N,\pv^N)$ (upper bubble) or
    coordinate-only dependence $\hat A(\rv^N)$ (lower bubble). In the
    latter case alternative forms involve the total local force
    density $\hat\Fv(\rv)$ or the splitting \eqref{EQybg} into its
    three constituent contributions.
    (b) Sum rules that result from specific observable choices.
    Global examples set $\hat A(\rv^N,\pv^N)$ as the density operator
    $\hat\rho(\rv)$, the interparticle force density operator
    $\hat\Fv_\rmint(\rv)$, and the total local force density operator
    $\hat \Fv(\rv)$, as corresponds to the sum
    rules~\eqref{EQrhoFextSumRule}, \eqref{EQFintFextSumRule}, and
    \eqref{EQFextFtotSumRule}, respectively.
    The local examples include the total external potential energy
    $\sum_i V_\rmext(\rv_i)$, the interparticle energy $u(\rv^N)$, and
    the sum of all particle coordinates $\sum_i\rv_i$, corresponding
    to the sum rules~\eqref{EQexternalEnergySumRule},
    \eqref{EQinternalEnergySumRule}, and
    \eqref{EQcenterOfMassSumRule}, respectively.
\label{FIG6}}
\end{figure}

In conclusion we have developed a statistical mechanical hyperforce
framework in generalization of the YBG equilibrium force balance
relationship \eqref{EQybg}. Our theory is based on previously
developed global \cite{hermann2021noether, hermann2022topicalReview,
  hermann2022variance} and local \cite{tschopp2022forceDFT,
  hermann2022quantum} shifting transformations on phase space. These
variable transformations leave the thermal physics invariant despite
an apparent dependence on the transformation parameter. The parameter
is a three-dimensional globally constant vector in case of global
symmetry, which applies to the entirety of the system, and a
position-dependent three-dimensional vector field for the locally
resolved case. Treating the corresponding phase space transformations
according to Noether's invariant variational calculus
\cite{noether1918} allows to systematically generate exact identities.

Here we have generalized this Noetherian concept to the equilibrium
average of an arbitrary given phase space function $\hat A$. The
resulting Noether identities couple in a specific manner the forces,
which the underlying Hamiltonian generates, to the observable $\hat A$
and its gradient with respect to the phase space variables. In the
position-resolved case we obtain localized correlation functions, with
the Dirac distribution generating microscopically sharp, but
statistically coarse-grained and hence well-accessible correlators.
In detail, we have presented the global hyperforce sum rule
\eqref{EQAepsGlobal} that applies to any given phase space function
$\hat A$. The local versions comprise Eqs.~\eqref{EQAepsLocal} and
\eqr{EQAepsCoordinatesOnly}, where the latter version applies to
momentum-independent observables. Decomposing the total force operator
into its ideal, interparticle, and external contributions leads to
\eqr{EQAepsCoordinatesExplicitForm}, which generalizes the equilibrium
force density balance \eqref{EQybg}.  An overview of these general
identities is shown in Fig.~\ref{FIG6}(a).

For a variety of relevant concrete choices of the form of~$\hat A$, we
have demonstrated explicitly their validity via carrying out many-body
simulations. This includes forces, energies, and entirely mechanical
quantities such as the center of mass; see Fig.~\ref{FIG6}(b) for a
summary of specific examples. We have shown that the sampling quality
and equilibration properties depend significantly on the type of
underlying sum rule. We argue that this behaviour forms a valuable
asset for systematic assessment of simulation quality.

Our hyperforce identities complement the virial~\cite{hansen2013},
hypervirial \cite{hirschfelder1960}, equipartition \cite{hansen2013}
and Yvon~\cite{hansen2013} theorems. Despite certain formal
similarities, we emphasize that the underlying phase space invariance
is more fundamental than derivations based on {\it ad hoc} partial
integration. Furthermore, the considered invariance operations
naturally lead to correlations with either global or locally resolved
forces, which are both simple to interpret and straightforward to
acquire in simulations.

We have shown how the global hyperforce identity~\eqref{EQAepsGlobal}
can alternatively be obtained from the Yvon theorem~\cite{hansen2013}.
Hence, as anticipated in the discussion by Rotenberg in
Ref.~\cite{rotenberg2020}, the Yvon theorem can indeed be a relevant
tool for force sampling.  However, the general localized Noether sum
rule \eqref{EQAepsLocal} reaches beyond the Yvon theorem in terms of
the momentum effects that are included. We have shown that the
derivation of the momentum-independent sum
rule~\eqref{EQAepsCoordinatesOnly} based on the Yvon theorem requires
to apply the ad hoc localized choice $\hat A \delta(\rv-\rv_i)$ and
summing over $i$. As a second step, treating the ideal contribution
explicitly allows to identify the sequence of emerging terms as the
one-body force operator $\hat\Fv(\rv)$ at any position. Conversely,
the Yvon theorem can be derived as a limit case from Noether
invariance upon shifting only one given particle $i$ according to
$\rv_i \to \rv_i + \eps_0$, and keeping unchanged all other particles
coordinates $\rv_j$ with $j \neq i$.

Besides the theoretical connections that the Noether hyperforce sum
rules establish, they can serve to carry out tests in theoretical and
simulation approaches, with possible fruitful connections to the
mapped-averaging force sampling framework \cite{purohit2019}. The
hyperforce sum rules can also provide a starting point, together with
the existing body of equilibrium sum rules \cite{upton1998, evans1990,
  henderson1984, henderson1985, triezenberg1972}, for the construction
of new inhomogeneous liquid state approximations.  We have exemplified
the use of sum rules in providing gauges for equilibration quality of
simulation data and we are confident in their future beneficial use in
machine-learning approaches such as the recent neural functional
theory \cite{sammueller2023neural, sammueller2023neuralTutorial}.

In the context of the use of machine-learning in Statistical Mechanics
\cite{delasheras2023perspective, sammueller2023neural,
  sammueller2023neuralTutorial, teixera2014, lin2019ml, lin2020ml,
  cats2022ml, yatsyshin2022} sum rules were shown to provide tests for
the successful construction of neural functionals both in
\cite{sammueller2023neural, sammueller2023neuralTutorial} and out of
equilibrium \cite{delasheras2023perspective}. These sum rules amount
to specific force properties, such as the vanishing of the global
interparticle force \cite{delasheras2023perspective} and the
interrelation of different orders of direct correlation functions
\cite{sammueller2023neural, sammueller2023neuralTutorial}. The present
much more general hyperforce framework can form much inspiration for
such approaches as well as for the recent force-based density
functional theory \cite{tschopp2022forceDFT, sammueller2022forceDFT},
which was compared \cite{sammueller2022forceDFT} to standard
fundamental measure theory \cite{roth2010}.

Furthermore, investigating hyperforce correlations in ionic systems,
see Refs.~\cite{minh2023faraday, minh2023noise, marbach2021reservoirs}
for recent work that addresses both concentration and charge
fluctuation behaviour, as are also relevant in confined systems
\cite{minh2023jcpConfined}, appears to be very promising.  This also
holds true for further interfacial physics \cite{upton1998, evans1990,
  henderson1984, henderson1985, triezenberg1972} and potentially for
the higher-order correlation functions of Refs.~\cite{zhang2020pnas,
  singh2023pnas, pihlajamaa2023}.

In our theoretical derivations we have relied on the grand canonical
ensemble, with fixed chemical potential $\mu$ and fluctuating number
of particles $N$. Carrying out formal manipulations in this way is
often more straightforward than working with fixed $N$, as is
appropriate for a canonical treatment. (Temperature is constant in
both ensembles.) A prominent example is to obtain the density profile
as a functional derivative $\rho(\rv)=\delta\Omega/\delta
V_\rmext(\rv)$ where crucially $\mu$ is kept fixed, rather than $N$,
upon building the functional derivative. This prototypical example
demonstrates the elegance of working grand canonically, and one could
expect that a similar situation applies for the thermal Noether
invariance. This, however, is not the case. Rather the phase space
shifting transformation, whether global by a constant $\eps_0$ or
locally resolved in position via a three-dimensional vector field
$\eps(\rv)$, is an entirely mechanical operation that applies equally
well canonically.  The shifting invariance gives a powerful new route
to correlation functions and and their sum rules, alternative to the
traditional method of integrating over degrees of freedom, as
pioneered by Yvon \cite{yvon1935} and Born and Green \cite{born1946}.

A detailed account of global shifting in the canonical ensemble is
provided in Ref.~\cite{hermann2022topicalReview}. The resulting
Noether force identities are analogous in form to the results from a
grand canonical treatment \cite{hermann2021noether}, with the sole
(and trivial) difference of the definition of the respective ensemble
averages. Here we find that the analogous situation holds for the
hyperforce identities. As they originate from phase space
transformations only, they are insensitive to the ensemble differences
between the canonical and the grand ensemble.  This theoretical fact
is corroborated by our computer simulation results, where we have
explicitly compared grand canonical Monte Carlo data and canonical
results, with the latter obtained via sampling under adaptive BD time
evolution \cite{sammueller2021}.

We have used overdamped Brownian time evolution as a means to sample
in thermal equilibrium. We find the adaptive Brownian dynamics time
stepping algorithm of Ref.~\cite{sammueller2021} to be a convenient
choice for our present purposes. The principle validity of the
hyperforce sum rules is nevertheless independent thereof and we expect
careful use of either the simpler Euler-Maruyama method
\cite{frenkel2023book, sammueller2021} or indeed Molecular Dynamics
\cite{frenkel2023book} to yield identical results. In our BD
simulations, we have sampled all correlation functions at equal time
as is the appropriate limit in equilibrium time evolution to recover
static thermal ensemble averages. As an aside,
Ref.~\cite{hermann2021noether} exploits Noether invariance in
nonequilibrium dynamics.

As a final note, we return to classical density functional theory and
its prowess in the description and incorporation of the fundamental
force correlators that emerge from the hyperforce concept. We recall
that classical density functional theory is based on a formally exact
variational principle which amounts to solving the following
Euler-Lagrange equation:
\begin{align}
  k_BT c_1(\rv,[\rho]) -V_\rmext(\rv)  &= k_BT\ln\rho(\rv) - \mu.
  \label{EQel}
\end{align}
Here $c_1(\rv,[\rho])$ is the one-body direct correlation function of
inhomogeneous liquid state theory. This is expressed as a density
functional via $c_1(\rv,[\rho])=-\beta \delta F_{\rm exc}[\rho] /
\delta \rho(\rv)$, where $F_{\rm exc}[\rho]$ is the intrinsic excess
Helmholtz free energy functional, which contains the interparticle
interactions, and $\delta/\delta\rho(\rv)$ denotes the functional
derivative with respect to the density profile. Solving \eqr{EQel} for
given $T, \mu$, and $V_\rmext(\rv)$ yields results for the equilibrium
density profile $\rho(\rv)$, which is hence the central variable of
density functional theory.

A multitude of connections with the current invariance theory emerge
naturally. From the density profile, using the hyperforce identities
one can obtain results for $\langle \hat \Fv(\rv)\sum_i\rv_i\rangle$
via \eqr{EQcenterOfMassSumRule}, for $\langle \hat \Fv_\rmext^0 \hat
\Fv(\rv)\rangle$ via \eqr{EQFextFtotSumRule}, for $\langle \hat
\Fv(\rv)\sum_i V_\rmext(\rv_i)\rangle$ via
\eqr{EQexternalEnergySumRule}, and upon building the gradient of the
density profile for $\langle \hat\Fv_\rmext^0\hat\rho(\rv)\rangle$ via
\eqr{EQrhoFextSumRule}.

We can make further progress by noting that within density functional
theory the interparticle force density is given by $\Fv_\rmint(\rv) =
k_BT \rho(\rv)\nabla c_1(\rv)$, where we have suppressed the
functional dependence on the density profile in the notation. That
this relationship holds can be seen from building the gradient of
\eqr{EQel} whereby $-\nabla\mu$ vanishes as the chemical potential is
constant, multiplying by $\rho(\rv)$, and comparing term-wise with the
force density balance relationship~\eqref{EQybg}.

Having obtained $\Fv_\rmint(\rv)$ in this (density functional) way
gives access to $\langle \hat \Fv(\rv) u(\rv^N)\rangle$ via
\eqr{EQinternalEnergySumRule}.  Building the gradient of the
interparticle force density via the product rule yields $\nabla
\Fv_\rmint(\rv) = k_BT [\nabla \rho(\rv)] \nabla c_1(\rv) + k_BT
\rho(\rv)\nabla\nabla c_1(\rv)$.  Again in principle the right hand
side is straightforward to evaluate in a typical numerical density
functional study as data for both $\rho(\rv)$ and $c_1(\rv)$ is
accessible. As a result the correlator $\langle \hat\Fv_\rmext^0 \hat
\Fv_\rmint(\rv)\rangle$ is available via the hyperforce sum rule
\eqref{EQFintFextSumRule}.

Hence standard results that are obtained within the density functional
framework allow to access a wealth of nontrivial force correlation
structure. This additional information is not redundant. We compare
with Evans and his coworkers' local compressibility
$\chi_\mu(\rv)=\partial \rho(\rv)/\partial \mu$, where similar to the
present force setup, one obtains $\chi_\mu(\rv)$ from processing the
density profile. In practice then analyzing $\chi_\mu(\rv)$ can shed
significantly more light on the physics than what is apparent from the
density profile alone, as has been demonstrated in a range of
insightful studies on drying at substrates and the important
phenomenon of hydrophobicity \cite{evans2015jpcm, evans2019pnas,
  coe2022prl}.

\bigskip

{\bf Code availability}. The simulation code to generate the data in
this study is available online at the following URL:
\href{https://gitlab.uni-bayreuth.de/bt306964/mbd/-/tree/hyperforce}
     {https://gitlab.uni-bayreuth.de/bt306964/mbd/-/tree/hyperforce}
\bigskip

{\bf Data availability.} The data is available upon reasonable
request.
\bigskip

{\bf Competing interests.} The authors declare no competing interests.
\bigskip

{\bf Author contributions.} SH and MS designed the research. All
authors carried out the analytical work. SR and FS performed the
numerical work. All authors wrote the paper.

\acknowledgments We thank Daniel de las Heras, Jonas K\"oglmayr, and
David Limmer for useful discussions.  This work is supported by the
German Research Foundation (DFG) via Project No.\ 436306241.


\end{document}